\documentstyle[epsfig,twocolumn,subfigure]{mn}

\title[Modelling clusters of galaxies by $f(R)$-gravity]{Modelling clusters of galaxies by $f(R)$-gravity}
\author[S. Capozziello, E. De Filippis, V. Salzano]{S. Capozziello, E. De Filippis, V. Salzano\\
Dipartimento di Scienze Fisiche, Universit\`{a} degli
Studi di Napoli Federico II and INFN, Sezione di Napoli, \\
Complesso Universitario di Monte S. Angelo, Via Cinthia, Edificio
N, 80126 Napoli, Italy\\}

\begin{document}

\date{Accepted xxx, Received yyy; in original form zzz}

\pagerange{\pageref{firstpage}--\pageref{lastpage}} \pubyear{0000}

\maketitle

\label{firstpage}

\begin{abstract}
 We consider the possibility that masses and gravitational
potentials of galaxy cluster, estimated at X-ray wavelengths,
could be explained without assuming huge amounts of dark matter,
but in the context of $f(R)$-gravity. Specifically, we take into
account the weak field limit  of such theories and show that the
corrected gravitational potential allows to  estimate the total
mass of a sample of $12$  clusters of galaxies. Results show that
such a gravitational potential provides a fair fit to the mass of
visible matter (i.e. gas + stars) estimated by X-ray observations,
without the need of additional dark matter while the size of the
clusters, as already observed at different scale for galaxies,
strictly depends on the interaction lengths of the corrections to
the Newtonian potential.
\end{abstract}


\section{Introduction}
Since the pioneering work by \cite{Zwi33}, the problems of high
mass-to-light ratios of galaxy clusters and of the rotation curves
of spiral galaxies  have been faced by asking for huge amounts of
unseen matter in the framework of Newtonian theory of gravity. It
is interesting to stress the fact that Zwicky addressed such an
issue dealing with {\it missing matter} and not with {\it dark
matter}.

Later on, several versions of the the cold dark matter model (CDM)
have been built starting from the assumption that a large amount
of non-baryonic matter (i.e. matter non-interacting with the
electromagnetic radiation) could account for the observations in
the framework of the standard Newtonian dynamics. Besides, by
adding a further ingredient, the cosmological constant $\Lambda$
\cite{CarLam,Sahni}, such a model (now $\Lambda$CDM) has become
the new cosmological  paradigm usually called the {\it concordance
model}.

In fact,  high quality data coming from the measurements of
cluster properties as the mass, the correlation function and the
evolution with redshift of their abundance
\cite{eke98,vnl02,bach03,bb03}, the Hubble diagram of Type Ia
Supernovae \cite{Riess04,ast05,clo05}, the optical surveys of
large scale structure \cite{pope04,cole05,eis05}, the anisotropies
of the cosmic microwave background \cite{Boom,WMAP}, the cosmic
shear measured from weak lensing surveys \cite{vW01,refr03} and
the Lyman\,-\,$\alpha$ forest absorption \cite{chd99,mcd04} are
evidences toward a spatially flat universe with a subcritical
matter content and undergoing a phase of accelerated expansion.
Interpreting all this information in a self-consistent model is
the main task of modern cosmology and $\Lambda$CDM model provides
a good fit to the most of the data \cite{Teg03,Sel04,sanch05} giving a
reliable snapshot of the today observed universe.

Nevertheless, it is affected by serious theoretical shortcomings
that have motivated the search for alternative candidates
generically referred to as {\it dark energy} or {\it
quintessence}. Such models  range from  scalar fields rolling down
self interaction potentials to phantom fields, from
phenomenological unified models of dark energy and dark matter to
alternative gravity theories  \cite{PB03,Pad03,tsu1}.

Essentially, dark energy (or any alternative component) has to act
as a negative pressure fluid which gives rise to an overall
acceleration of the Hubble fluid. Despite of the clear mechanisms
generating the observed cosmological dynamics, the nature and the fundamental
properties of  dark energy remain essentially unknown
notwithstanding the great theoretical efforts made up to now.

The
situation for dark matter is similar: its clustering and
distribution properties are fairly well known at every scale but
its nature is unknown, up to now, at a fundamental level.

On the other hand, the need of  unknown components as dark energy
and dark matter could be considered nothing else but as a signal
of the breakdown of Einstein General Relativity at astrophysical
(galactic and extragalactic) and cosmological scales.

In this context, Extended Theories of Gravity could be, in
principle, an interesting alternative to explain cosmic
acceleration without any dark energy and large scale structure
without any dark matter. In their simplest version, the Ricci
curvature scalar $R$, linear in the Hilbert-Einstein action, could
be replaced by a generic function $f(R)$ whose true form could be
"reconstructed" by the data. In fact, there is no a priori reason
to consider the gravitational Lagrangian linear in the Ricci
scalar while observations and experiments could contribute to
define and constrain the "true" theory of gravity. For a
discussion on this topic, see \cite{capfra,odi,faraoni}.

In a cosmological framework, such an approach lead to modified
Friedmann equations that can be formally written in the usual form
by defining an effective  {\it curvature fluid} giving rise to a negative pressure
which drives the cosmic acceleration
\cite{capozzcurv,noirev,cdtt,capfra}. Also referred to as
$f(R)$ theories, this approach is recently  extensively studied both
from the theoretical  and the observational point of view (see
e.g.  \cite{noiijmpd,ccf,borowiec}). Moreover, this same approach
has been also proposed, in early cosmology, as a mechanism giving
rise to an inflationary era without the need of any inflaton field
\cite{Star80}.

All this amount of work has been, essentially,  concentrated on
the cosmological applications  and have convincingly demonstrated
that extended theories of gravity (in particular $f(R)$ gravity)
are indeed able to explain the cosmic speed up and fairly fit the
available data-sets and, hence, represents a viable alternative to
the dark energy models \cite{cct}.

Changing the gravity sector could have consequences not only at
cosmological scales, but also at the galactic and cluster scales
so that it is mandatory to investigate the low energy limit of
such theories. A strong debate is  open with different  results
arguing in favor
\cite{dick,sotiriou,cembranos,navarro,allrugg,ppnantro} or against
\cite{dolgov,chiba,olmo} such models. It is worth noting that, as
a general result, higher order theories of gravity cause the
gravitational potential to deviate from its Newtonian $1/r$
scaling \cite{b10,hj,hjrev,cb05,sobouti,Mendoza} even if such
deviations may  be  vanishing.

In  \cite{CapCardTro07},  the Newtonian limit of power law $f(R) =
f_0 R^n$ theories has been investigated, assuming that the metric
in the low energy limit ($\Phi/c^2 << 1$) may be taken as
Schwarzschild\,-\,like. It turns out that a power law term
$(r/r_c)^{\beta}$ has to be added to the Newtonian $1/r$ term in
order to get the correct gravitational potential. While the
parameter $\beta$ may be expressed analytically as a function of
the slope $n$ of the $f(R)$ theory, $r_c$ sets the scale where the
correction term starts being significant. A particular range of
values of $n$  has been investigated so that the corrective term
is an increasing function of the radius $r$ thus causing an
increase of the rotation curve with respect to the Newtonian one
and offering the possibility to fit the galaxy rotation curves
without the need of further dark matter components.

A set of low surface brightness (LSB) galaxies with extended and
well measured rotation curves has been considered
\cite{dbb02,db05}. These systems are supposed to be dark matter
dominated, and successfully fitting data without dark matter is a
strong evidence in favor of the approach (see also \cite{Frigerio}
for an independent analysis using another sample of galaxies).
Combined with the hints coming from the cosmological applications,
one should have, in principle, the possibility to address both the
dark energy and dark matter problems resorting to the same well
motivated fundamental theory \cite{prl,Koivisto,lobo}.
Nevertheless, the simple power law $f(R)$ gravity is nothing else
but a toy-model which fail if one tries to achieve a comprehensive
model for all the cosmological dynamics, ranging from the early
universe, to the large scale structure up to the late accelerated
era \cite{prl,Koivisto}.

A fundamental issue is related to  clusters and  superclusters of
galaxies. Such structures, essentially, rule the large scale
structure, and are the intermediate step between galaxies and
cosmology. As the galaxies, they appear dark-matter dominated but
the distribution  of dark matter component seems clustered and
organized in a very different way with respect to galaxies. It
seems that dark matter is ruled by the scale and also its
fundamental nature could depend on the scale. For a comprehensive
review see \cite{Bah96}.

In the philosophy of Extended Theories of Gravity, the issue is
now to reconstruct the mass profile of clusters {\it without} dark
matter, i.e. to find out corrections to the Newton potential
producing the same dynamics as dark matter  but starting from a
well motivated theory. This is the goal of this paper.

As we will see, the problem is very different with respect to that
of galaxies and we  need different  corrections to the
gravitational potential in order to consistently fit the cluster
mass profiles.

The paper is organized as follows. In
Section~\ref{sec:ext_theories}, also referring to other results,
 we discuss the weak field limit
of $f(R)$-gravity showing that an $f(R)$ theory is, in principle, quite
 fair to address the cluster problem \cite{lobo}. In Section~\ref{sec:ext_systems},
starting from the corrected potential previously derived for a
point-like masses,  we generalize the results  to  extended
spherically symmetric systems  as required for well shaped cluster
models of the Bautz and Morgan classification \cite{Bau70}.
Section~\ref{sec:mass_profiles} is devoted to the description of the
general properties of galaxy clusters. In
Section~\ref{sec:sample}, the sample of galaxy clusters which we
are going to fit is discussed in details. Results are presented in
Section~\ref{sec:results}, while Section~\ref{sec:discussion} is
devoted to the discussion and the conclusions.

\section{$f(R)$-gravity}
\label{sec:ext_theories}

Let us consider the general action\,:

\begin{equation}\label{actfR}
{\cal A}\, = \,\int
d^4x\sqrt{-g}\left[f(R)+{\cal X}{\cal L}_m\right]\,,
\end{equation}
where $f(R)$ is an  analytic function of the Ricci scalar $R$, $g$ is the
determinant of the metric $g_{\mu\nu}$, ${\displaystyle {\cal X}=\frac{16\pi
G}{c^4}}$ is the coupling constant and ${\cal L}_m$ is
the standard perfect-fluid matter Lagrangian. Such an
action is the straightforward generalization of the
Hilbert-Einstein action of GR obtained for $f(R)=R$.
Since we are considering the  metric approach, field equations
are obtained by varying (\ref{actfR}) with respect to the
metric\,:

\begin{eqnarray}\label{fe}
f'R_{\mu\nu}-\frac{1}{2}fg_{\mu\nu}-f'_{;\mu\nu}+g_{\mu\nu}\Box
f'=\frac{\mathcal{X}}{2}T_{\mu\nu}\,.
\end{eqnarray}
where $T_{\mu\nu}=\frac{-2}{\sqrt{-g}}\frac{\delta(\sqrt{-g}{\cal
L}_m)}{\delta g^{\mu\nu}}$ is the energy momentum tensor of
matter, the prime indicates the derivative with respect to $R$ and
$\Box={{}_{;\sigma}}^{;\sigma}$. We adopt the signature
$(+,-,-,-)$.

As discussed in details in  \cite{noi-prd}, we deal with the
Newtonian and the post-Newtonian limit of $f(R)$ - gravity on a
spherically symmetric background. Solutions for the field
equations can be obtained by imposing the spherical symmetry
\cite{arturocqg}:

\begin{eqnarray}\label{me}
ds^2\,=\,g_{00}(x^0,r)d{x^0}^2+g_{rr}(x^0,r)dr^2-r^2d\Omega
\end{eqnarray}
where $x^0\,=\,ct$ and $d\Omega$ is the angular element.

To develop the  post-Newtonian limit of the theory,
one can consider a perturbed metric with
respect to a Minkowski background
$g_{\mu\nu}\,=\,\eta_{\mu\nu}+h_{\mu\nu}$. The metric coefficients
can be developed as:

\begin{eqnarray}\label{definexpans}
\left\{\begin{array}{ll} g_{tt}(t,
r)\simeq1+g^{(2)}_{tt}(t,r)+g^{(4)}_{tt}(t,r)
\\\\
g_{rr}(t,r)\simeq-1+g^{(2)}_{rr}(t,r)\\\\
g_{\theta\theta}(t,r)=-r^2\\\\
g_{\phi\phi}(t,r)=-r^2\sin^2\theta
\end{array}\right.\,,
\end{eqnarray}
where we put, for the sake of simplicity, $c\,=\,1$\,,\
$x^0=ct\rightarrow t$.
We want to obtain the most general result without imposing particular forms for the $f(R)$-Lagrangian.
We only consider analytic Taylor expandable functions
\begin{eqnarray}\label{sertay}
f(R)\simeq
f_0+f_1R+f_2R^2+f_3R^3+...\,.
\end{eqnarray}
To obtain the post-Newtonian approximation of $f(R)$ - gravity, one
has to plug the expansions (\ref{definexpans}) and
(\ref{sertay}) into the field equations (\ref{fe})
and then expand the system up to the orders $O(0),\, O(2)$ and $O(4)$ . This
approach provides general results and specific (analytic)
Lagrangians are selected by  the coefficients $f_i$ in
(\ref{sertay}) \cite{noi-prd}.

If we now consider the $O(2)$ - order of approximation, the field equations
(\ref{fe}),  in the vacuum case, results to be

\begin{eqnarray}\label{eq2}
\left\{\begin{array}{ll}
f_1rR^{(2)}-2f_1g^{(2)}_{tt,r}+8f_2R^{(2)}_{,r}-f_1rg^{(2)}_{tt,rr}+4f_2rR^{(2)}=0
\\\\
f_1rR^{(2)}-2f_1g^{(2)}_{rr,r}+8f_2R^{(2)}_{,r}-f_1rg^{(2)}_{tt,rr}=0
\\\\
2f_1g^{(2)}_{rr}-r[f_1rR^{(2)}
\\\\
-f_1g^{(2)}_{tt,r}-f_1g^{(2)}_{rr,r}+4f_2R^{(2)}_{,r}+4f_2rR^{(2)}_{,rr}]=0
\\\\
f_1rR^{(2)}+6f_2[2R^{(2)}_{,r}+rR^{(2)}_{,rr}]=0
\\\\
2g^{(2)}_{rr}+r[2g^{(2)}_{tt,r}-rR^{(2)}+2g^{(2)}_{rr,r}+rg^{(2)}_{tt,rr}]=0
\end{array} \right.\end{eqnarray}
It is evident that the trace equation (the fourth in the system
(\ref{eq2})), provides a differential equation with respect to the
Ricci scalar which allows to solve  the system  at $O(2)$
- order. One obtains the general solution\,:
\begin{eqnarray}\label{sol}
\left\{\begin{array}{ll}
g^{(2)}_{tt}=\delta_0-\frac{2GM}{f_1r}-\frac{\delta_1(t)e^{-r\sqrt{-\xi}}}{3\xi
r}+\frac{\delta_2(t)e^{r\sqrt{-\xi}}}{6({-\xi)}^{3/2}r}
\\\\
g^{(2)}_{rr}=-\frac{2GM}{f_1r}+\frac{\delta_1(t)[r\sqrt{-\xi}+1]e^{-r\sqrt{-\xi}}}{3\xi
r}-\frac{\delta_2(t)[\xi r+\sqrt{-\xi}]e^{r\sqrt{-\xi}}}{6\xi^2r}
\\\\
R^{(2)}=\frac{\delta_1(t)e^{-r\sqrt{-\xi}}}{r}-\frac{\delta_2(t)\sqrt{-\xi}e^{r\sqrt{-\xi}}}{2\xi
r}\end{array} \right.
\end{eqnarray}
where $\xi\doteq\displaystyle\frac{f_1}{6f_2}$, $f_1$ and $f_2$
are the expansion coefficients obtained by the $f(R)$-Taylor series.
 In the limit $f\rightarrow R$, for a point-like source of mass $M$ we recover the standard
Schwarzschild solution. Let us notice that
the integration constant $\delta_0$ is  dimensionless,
while the two arbitrary time-functions  $\delta_1(t)$ and
$\delta_2(t)$ have respectively the dimensions of $lenght^{-1}$
and $lenght^{-2}$; $\xi$ has the dimension $lenght^{-2}$. As extensively discussed in \cite{noi-prd},
the functions $\delta_i(t)$ ($i=1,2$)
are completely arbitrary since the differential equation system
(\ref{eq2}) depends only on spatial derivatives. Besides, the integration constant $\delta_0$
can be set to zero, as in the standard theory of  potential,  since it
represents an unessential additive quantity.
 In order to obtain the physical prescription of the asymptotic flatness
at infinity, we can discard
the Yukawa growing mode in (\ref{sol}) and then the metric is \,:

\begin{eqnarray}\label{mesol}
ds^2&=&\biggl[1-\frac{2GM}{f_1r}-\frac{\delta_1(t)e^{-r\sqrt{-\xi}}}{3\xi
r}\biggr]dt^2\nonumber\\
&-&\biggl[1+\frac{2GM}{f_1r}-\frac{\delta_1(t)(r\sqrt{-\xi}+1)e^{-r\sqrt{-\xi}}}{3\xi
r}\biggr]dr^2\nonumber\\
&-&r^2d\Omega\,.
\end{eqnarray}
The Ricci scalar curvature is
\begin{equation}
R\,=\,\frac{\delta_1(t)e^{-r\sqrt{-\xi}}}{r}\,.
\end{equation}

The solution can be given also in terms of
gravitational potential. In particular, we have an
explicit Newtonian-like term into the definition. The first of (\ref{sol}) provides the
second order solution in term of the metric expansion (see the
definition (\ref{definexpans})). In particular,
it is $g_{tt}\,=\,1+2\phi_{grav}\,=\,1+g_{tt}^{(2)}$ and then the
gravitational potential of an analytic $f(R)$-theory
is

\begin{eqnarray}\label{gravpot}
\phi_{grav}\,=\,-\frac{GM}{f_1r}-\frac{\delta_1(t)e^{-r\sqrt{-\xi}}}{6\xi
r}\,.
\end{eqnarray}
Among the possible analytic $f(R)$-models, let us consider the Taylor expansion where the cosmological term (the above $f_0$) and terms higher than second have been discarded. We rewrite the Lagrangian (\ref{sertay}) as
\begin{equation}
f(R) \sim a_1 R + a_2 R^2 + ...
\end{equation}
and specify the above gravitational potential (\ref{gravpot}), generated
by a point-like matter distribution, as:
\begin{equation}
\label{gravpot1} \phi(r) = -\frac{3 G M}{4 a_1
r}\left(1+\frac{1}{3}e^{-\frac{r}{L}}\right)\,,
\end{equation}
where
\begin{equation}\label{lengths model}
L \equiv L(a_{1},a_{2}) =  \left( -\frac{6 a_2}{a_1}
\right)^{1/2}\,.
\end{equation}
$L$ can be defined as  the {\it interaction length} of the
problem\footnote{Such a length is
function of the series coefficients, $a_{1}$
and $a_{2}$, and it is not a free independent parameter in the following fit
procedure.} due to the correction to the Newtonian potential.
We have changed the notation to remark that we are doing only a
specific choice in the wide class of potentials (\ref{gravpot}),
but the following considerations are completely general.

\section{Extended systems}
\label{sec:ext_systems}

The gravitational potential (\ref{gravpot1}) is a
point-like one. Now we have to generalize this solution to
extended systems. Let us describe galaxy clusters as
spherically symmetric systems and then we have to extend the above considerations
to this geometrical configuration. We simply consider
the system composed by many infinitesimal mass elements $dm$ each
one contributing with a point-like gravitational potential. Then,
summing up all terms, namely integrating them on a spherical
volume, we obtain a suitable potential. Specifically, we have to solve
 the integral:
\begin{equation}
\Phi(r) = \int_{0}^{\infty} r'^{2} dr' \int_{0}^{\pi} \sin \theta'
d\theta' \int_{0}^{2\pi} d\omega' \hspace{0.1cm} \phi(r')\,.
\end{equation}
The point-like potential (\ref{gravpot1})can be split in two terms. The
\textit{Newtonian} component is
\begin{equation}
\phi_{N}(r) = -\frac{3 G M}{4 a_1 r}
\end{equation}
The extended integral of such a part is  well-known (apart from the
numerical constant $\frac{3}{4 a_{1}}$) expression. It is
\begin{equation}
\Phi_{N}(r) = -\frac{3}{4 a_1}\frac{G M(<r)}{r} \\
\end{equation}
where $M(<r)$ is the mass enclosed in a sphere with radius $r$.
The \textit{correction} term
\begin{equation}
\phi_{C}(r) = -\frac{G M}{4 a_1}\frac{e^{-\frac{r}{L}}}{r}
\end{equation}
considering some analytical steps in the integration of the angular part,
give the expression
\begin{equation}
\Phi_{C}(r) = - \frac{2 \pi G}{4} \cdot L \int_{0}^{\infty} dr' r'
\rho(r') \cdot \frac{e^{-\frac{|r-r'|}{L}} -
e^{-\frac{|r+r'|}{L}}}{r}
\end{equation}
The radial integral is  numerically estimated once
the mass density is given. We underline a fundamental difference
between such a term and the Newtonian one: while in the latter, the
matter outside the spherical shell of radius $r$ does not
contribute to the potential, in the former, external matter takes
part to the integration procedure. For this reason we split the corrective
potential in two terms:
\begin{itemize}
    \item if $r' < r$:
      {\setlength\arraycolsep{0.2pt}
      \begin{eqnarray}
      \Phi_{C,int}(r) &=& - \frac{2 \pi G}{4} \cdot L \int_{0}^{r} dr' r'
      \rho(r') \cdot \frac{e^{-\frac{|r-r'|}{L}} -
      e^{-\frac{|r+r'|}{L}}}{r}  \nonumber \\
     & =& - \frac{2 \pi G}{4} \cdot L \int_{0}^{r} dr' r' \rho(r') \cdot
      e^{-\frac{r+r'}{L}} \left( \frac{-1 + e^{\frac{2r'}{L}}}{r}
      \right)\nonumber
      \end{eqnarray}}
   \item if $r' > r$:
      {\setlength\arraycolsep{0.2pt}
      \begin{eqnarray}
      \Phi_{C,ext}(r) &=& - \frac{2 \pi G}{4} \cdot L \int_{r}^{\infty}
      dr' r' \rho(r') \cdot \frac{e^{-\frac{|r-r'|}{L}} -
      e^{-\frac{|r+r'|}{L}}}{r} = \nonumber \\
      &=& - \frac{2 \pi G}{4} \cdot L \int_{r}^{\infty} dr' r' \rho(r')
      \cdot e^{-\frac{r+r'}{L}} \left( \frac{-1 + e^{\frac{2r}{L}}}{r}
      \right)\nonumber
      \end{eqnarray}}
\end{itemize}
The total potential of the spherical mass distribution will be
\begin{equation}\label{eq:total corrected potential}
\Phi(r) = \Phi_{N}(r) + \Phi_{C,int}(r) + \Phi_{C,ext}(r)
\end{equation}
As we will show below, for our purpose, we need
the gravitational potential derivative with respect to the variable $r$; the
two derivatives may not be evaluated analytically so we estimate
them numerically, once we have given an expression for the
\textit{total} mass density $\rho(r)$. While the Newtonian
term gives the simple expression:
\begin{equation}
- \frac{\mathrm{d}\Phi_{N}}{\mathrm{d}r}(r) = - \frac{3}{4 a_{1}}
\frac{G M(<r)}{r^{2}}
\end{equation}
the internal and external derivatives of the corrective
potential terms are more involved. We do not give them explicitly for
sake of brevity, but  they are
integral-functions of the form
\begin{equation}
{\mathcal{F}}(r, r') = \int_{\alpha(r)}^{\beta(r)} dr' \ f(r,r')
\end{equation}
from which one has:
{\setlength\arraycolsep{0.2pt}
\begin{eqnarray}
\frac{{\mathrm{d}}{\mathcal{F}}(r, r')}{\mathrm{d}r} &=&
\int_{\alpha(r)}^{\beta(r)} dr'
\frac{{\mathrm{d}}f(r,r')}{{\mathrm{d}}r} + \nonumber \\
&-& f(r,\alpha(r)) \frac{{\mathrm{d}}\alpha}{{\mathrm{d}}r}(r) +
f(r,\beta(r)) \frac{{\mathrm{d}}\beta}{{\mathrm{d}}r}(r)
\end{eqnarray}}
Such an expression is numerically derived once the integration extremes are given.
A general consideration is in order at this point. Clearly, the Gauss theorem holds only for the Newtonian part  since, for this term, the force law scales as $1/r^2$. For the total potential (\ref{gravpot1}), it does not hold anymore due to the correction. From a physical point of view, this is not a problem because the full conservation laws are determined, for $f(R)$-gravity,  by the contracted Bianchi identities  which assure the self-consistency. For a detailed discussion, see \cite{CapCardTro07,capfra,faraoni}.

\section{Cluster mass profiles}
\label{sec:mass_profiles}

Clusters of galaxies are generally considered self-bound gravitational systems
with spherical symmetry and in hydrostatic equilibrium if virialized. The last
two hypothesis are still widely used, despite of the fact that it has
been widely proved that most clusters show more complex
morphologies and/or signs of strong interactions or dynamical
activity, especially in their innermost regions \cite{Chak08,DeFil05}. \\
Under the hypothesis of  spherical symmetry in hydrostatic
equilibrium, the structure equation can be derived from the
collisionless Boltzmann equation
\begin{equation}\label{Boltzmann equation}
\frac{d}{dr}(\rho_{gas}(r) \hspace{0.1cm} \sigma^{2}_{r}) +
\frac{2\rho_{gas}(r)}{r}(\sigma^{2}_{r}-\sigma^{2}_{\theta,\omega})
= -\rho_{gas}(r) \cdot \frac{d\Phi(r)}{dr}
\end{equation}
where $\Phi$ is the gravitational potential of the cluster,
$\sigma_{r}$ and $\sigma_{\theta,\omega}$ are the mass-weighted
velocity dispersions in the radial and tangential directions,
respectively, and $\rho$ is gas mass-density. For an isotropic
system, it is
\begin{equation}\label{velocity dispersion}
\sigma_{r} = \sigma_{\theta,\omega}
\end{equation}
The pressure profile can be related to these quantities by
\begin{equation}\label{pressure}
P(r) = \sigma^{2}_{r} \rho_{gas}(r)
\end{equation}
Substituting Eqs. (\ref{velocity dispersion}) and (\ref{pressure})
into Eq. (\ref{Boltzmann equation}), we have, for an isotropic sphere,
\begin{equation}\label{isotropic sphere}
\frac{d P(r)}{dr} = - \rho_{gas}(r) \frac{d\Phi(r)}{dr}
\end{equation}
For a gas sphere with temperature profile $T(r)$, the velocity
dispersion becomes
\begin{equation}\label{temperature}
\sigma^{2}_{r} = \frac{k T(r)}{\mu m_{p}}
\end{equation}
where $k$ is the Boltzmann constant, $\mu \approx 0.609$ is the
mean mass particle and $m_{p}$ is the proton mass. Substituting
Eqs. (\ref{pressure}) and (\ref{temperature}) into Eq.
(\ref{isotropic sphere}), we obtain
\[
\frac{d}{dr} \left( \frac{k T(r)}{\mu m_{p}} \rho_{gas}(r) \right)
= -\rho_{gas}(r) \frac{d \Phi}{dr}
\]
or, equivalently,
\begin{equation}\label{eq:Boltzmann potential}
-\frac{d\Phi}{dr} = \frac{k T(r)}{\mu m_{p}
r}\left[\frac{d\ln\rho_{gas}(r)}{d\ln r} + \frac{d\ln T(r)}{d\ln
r}\right]
\end{equation}
Now the total gravitational potential of the cluster is:
\begin{equation}\label{eq:total corrected potential1}
\Phi(r) = \Phi_{N}(r) + \Phi_{C}(r)
\end{equation}
with
\begin{equation}
\Phi_{C}(r) = \Phi_{C,int}(r) + \Phi_{C,ext}(r)
\end{equation}
It is worth underlining that if we consider \textit{only} the
standard Newtonian  potential, the \textit{total}
cluster mass $M_{cl,N}(r)$ is composed by gas mass $+$
mass of galaxies $+$ cD-galaxy mass $+$ dark matter and it is given by the
expression:
{\setlength\arraycolsep{0.2pt}
\begin{eqnarray}
\label{eq:M_tot} M_{cl,N}(r) &=& M_{gas}(r) + M_{gal}(r) +
M_{CDgal}(r) + M_{DM}(r)  \nonumber
\\
&=& - \frac{k T(r)}{\mu m_{p} G} r
\left[\frac{d\ln\rho_{gas}(r)}{d\ln r}+\frac{d\ln T(r)}{d\ln
r}\right]
\end{eqnarray}}
$M_{cl,N}$  means the standard estimated \textit{Newtonian} mass.
Generally the galaxy part contribution  is considered
negligible with respect to the other two components so we have:
\[
M_{cl,N}(r) \approx M_{gas}(r) + M_{DM}(r) \approx
\]
\[
\hspace{1.35cm} \approx - \frac{k T(r)}{\mu m_{p}} r
\left[\frac{d\ln\rho_{gas}(r)}{d\ln r}+\frac{d\ln T(r)}{d\ln
r}\right]
\]
Since the gas-mass estimates are provided by X-ray observations, the
equilibrium equation can be used to derive the amount of dark
matter present in a cluster of galaxies and its spatial
distribution.

Inserting the previously defined \textit{extended-corrected}
potential of Eq.~(\ref{eq:total corrected potential1}) into
Eq.~(\ref{eq:Boltzmann potential}), we obtain:
\begin{equation}
\label{eq:corrected_mass} -\frac{\mathrm{d}\Phi_{N}}{\mathrm{d}r}
-\frac{\mathrm{d}\Phi_{C}}{\mathrm{d}r} =\frac{k T(r)}{\mu m_{p}
r}\left[\frac{\mathrm{d}\ln\rho_{gas}(r)}{\mathrm{d}\ln r} +
\frac{\mathrm{d}\ln T(r)}{\mathrm{d}\ln r}\right]
\end{equation}
from which the \textit{extended-corrected} mass estimate follows:
{\setlength\arraycolsep{0.2pt}
\begin{eqnarray}\label{eq:fit relation}
M_{cl,EC}(r) &+& \frac{4 a_{1}}{3G} r^{2}
\frac{\mathrm{d}\Phi_{C}}{\mathrm{d}r}(r) = \nonumber \\ &=&
\frac{4 a_{1}}{3} \left[ - \frac{k T(r)}{\mu m_{p}G} r
\left(\frac{d\ln\rho_{gas}(r)}{d\ln r}+\frac{d\ln T(r)}{d\ln
r}\right) \right]
\end{eqnarray}}
Since the use of a corrected potential avoids, in principle, the additional
requirement of dark matter, the total cluster mass, in this case,
is given by:
\begin{equation}
M_{cl,EC}(r) = M_{gas}(r) + M_{gal}(r) + M_{CDgal}(r)
\end{equation}
and the mass density in the $\Phi_{C}$ term is
\begin{equation}
\rho_{cl,EC}(r) = \rho_{gas}(r) + \rho_{gal}(r) + \rho_{CDgal}(r)
\end{equation}
with the  density components derived from observations.

In this work, we will use  Eq.~(\ref{eq:fit relation}) to
compare the baryonic mass profile $M_{cl,EC}(r)$, estimated from
observations, with the theoretical deviation from the Newtonian
gravitational potential, given by the expression ${\displaystyle -\frac{4
a_{1}}{3G} r^{2} \frac{\mathrm{d}\Phi_{C}}{\mathrm{d}r}(r)}$.
Our goal is to reproduce the observed mass profiles for
a sample of galaxy clusters.

\section{Galaxy Cluster Sample}
\label{sec:sample}

The formalism described in \S~\ref{sec:mass_profiles} can be applied
to a sample of $12$ galaxy clusters. We shall use the cluster sample studied in
~\cite{Vik05,Vik06} which consists of $13$ low-redshift clusters
spanning a temperature range $0.7\div 9.0\ {\rm keV}$ derived from
high quality {\it Chandra} archival data. In all these clusters, the surface
brightness and the gas temperature profiles are measured out to large
radii, so that mass estimates can be extended up to r$_{500}$ or
beyond.

\subsection{Gas Density Model}
\label{sec:gas_model}

The gas density distribution of the clusters in the
sample is described by the analytic model proposed in~\cite{Vik06}. Such a model
modifies the classical $\beta-$model to represent the characteristic
properties of the observed X-ray surface brightness profiles, i.e.
the power-law-type cusps of gas density in the cluster center,
instead of a flat core and the steepening of the
brightness profiles at large radii. Eventually, a second $\beta-$model,
with a small core radius, is added to improve  the model
close to the cluster cores. The  analytical
form for the particle emission  is given by:
{\setlength\arraycolsep{0.2pt}
\begin{eqnarray}
\label{gas density vik} n_{p}n_{e} = n_{0}^{2} \cdot
\frac{(r/r_{c})^{-\alpha}}{(1+r^{2}/r_{c}^{2})^{3\beta-\alpha/2}}
& \cdot&\frac{1}{(1+r^{\gamma}/r_{s}^{\gamma})^{\epsilon/\gamma}}+
\nonumber \\
&+& \frac{n_{02}^{2}}{(1+r^{2}/r_{c2}^{2})^{3\beta_{2}}}
\end{eqnarray}}
which can be easily converted to a mass density using the relation:
\begin{equation}
\label{eq:gas_density} \rho_{gas} = n_T \cdot \mu m_p =
\frac{1.4}{1.2} n_e m_p
\end{equation}
where $n_T$ is the total number density of particles in the gas.
The resulting model has a large number of parameters, some of
which do not have a direct physical interpretation. While this can often
be inappropriate and computationally inconvenient, it suits well
our case, where the main requirement is a detailed qualitative
description of the cluster profiles.\\
In \cite{Vik06},  Eq.~(\ref{gas density vik}) is applied to a restricted
range of distances from the cluster center, i.e. between an inner
cutoff $r_{min}$, chosen to exclude the central temperature bin
($\approx 10\div 20\ {\rm kpc}$) where the ICM is likely to be
multi-phase, and $r_{det}$, where the X-ray surface brightness is at
least $3 \sigma$ significant. We have extrapolated the above
function to values outside this restricted range using the
following criteria:
\begin{itemize}
  \item for $r < r_{min}$, we have performed a linear extrapolation
  of the first three terms out to $r = 0$ kpc;
  \item for $r > r_{det}$, we have performed a linear extrapolation
  of the last three terms out to a distance $\bar{r}$ for which
  $\rho_{gas}(\bar{r})=\rho_{c}$, $\rho_{c}$ being the critical
  density of the Universe at the cluster redshift:
  $\rho_{c} = \rho_{c,0} \cdot (1 + z)^{3}$. For radii larger than $\bar{r}$,
  the gas density is assumed constant at $\rho_{gas}(\bar{r})$.
\end{itemize}
We point out that, in Table~\ref{tab1}, the radius limit $r_{min}$
is almost the same as given in the previous definition. When the
value given by \cite{Vik06} is less than the cD-galaxy radius, which is
defined in the next section, we  choose  this last one as the lower
limit. On the contrary, $r_{max}$ is quite different from
$r_{det}$: it is fixed by considering the higher value of temperature profile
and not by imaging methods. \\
We then compute the gas mass $M_{gas}(r)$ and the total mass
$M_{cl,N}(r)$, respectively, for all clusters in our sample,
substituting Eq.~(\ref{gas density vik}) into
Eqs.~(\ref{eq:gas_density}) and (\ref{eq:M_tot}), respectively;
the gas temperature profile {\bf is} described in details in
\S~\ref{sec:T_prof}. The resulting mass values, estimated at
$r=r_{max}$, are listed in Table~\ref{tab1}.

\begin{table*}
 \begin{minipage}{142mm}
 \caption{Column 1: Cluster name. Column2: Richness. Column 3: cluster total mass.
 Column 4: gas mass. Column 5: galaxy mass. Column 6: cD-galaxy mass.  Gas and total mass values are estimated at $r=r_{max}$. Column 7:
 ratio of total galaxy mass to gas mass. Column 8: minimum radius. Column 9: maximum
radius.}
 \label{tab1}
 \begin{tabular}{|c|c|c|c|c|c|c|c|c|}
    \hline
  name & R &$M_{cl,N}$ & $M_{gas}$ & $M_{gal}$  & $M_{cDgal}$  & $\frac{gal}{gas}$ & $r_{min}$ & $r_{max}$  \\
  &  & $(M_{\odot})$ &($M_{\odot}$)& ($M_{\odot}$) & ($M_{\odot}$)  &  & (kpc) & (kpc) \\
  \hline
  \hline
  A133    & 0 & $4.35874\cdot10^{14}$ & $2.73866\cdot10^{13}$ & $5.20269\cdot10^{12}$ & $1.10568\cdot10^{12}$ & $0.23$ &  $86$ & $1060$ \\
  A262    & 0 & $4.45081\cdot10^{13}$ & $2.76659\cdot10^{12}$ & $1.71305\cdot10^{11}$ & $5.16382\cdot10^{12}$ & $0.25$ &  $61$ & $ 316$ \\
  A383    & 2 & $2.79785\cdot10^{14}$ & $2.82467\cdot10^{13}$ & $5.88048\cdot10^{12}$ & $1.09217\cdot10^{12}$ & $0.25$ &  $52$ & $ 751$ \\
  A478    & 2 & $8.51832\cdot10^{14}$ & $1.05583\cdot10^{14}$ & $2.15567\cdot10^{13}$ & $1.67513\cdot10^{12}$ & $0.22$ &  $59$ & $1580$ \\
  A907    & 1 & $4.87657\cdot10^{14}$ & $6.38070\cdot10^{13}$ & $1.34129\cdot10^{13}$ & $1.66533\cdot10^{12}$ & $0.24$ & $563$ & $1226$ \\
  A1413   & 3 & $1.09598\cdot10^{15}$ & $9.32466\cdot10^{13}$ & $2.30728\cdot10^{13}$ & $1.67345\cdot10^{12}$ & $0.26$ &  $57$ & $1506$ \\
  A1795   & 2 & $5.44761\cdot10^{14}$ & $5.56245\cdot10^{13}$ & $4.23211\cdot10^{12}$ & $1.93957\cdot10^{12}$ & $0.11$ &  $79$ & $1151$ \\
  A1991   & 1 & $1.24313\cdot10^{14}$ & $1.00530\cdot10^{13}$ & $1.24608\cdot10^{12}$ & $1.08241\cdot10^{12}$ & $0.23$ &  $55$ & $ 618$ \\
  A2029   & 2 & $8.92392\cdot10^{14}$ & $1.24129\cdot10^{14}$ & $3.21543\cdot10^{13}$ & $1.11921\cdot10^{12}$ & $0.27$ &  $62$ & $1771$ \\
  A2390   & 1 & $2.09710\cdot10^{15}$ & $2.15726\cdot10^{14}$ & $4.91580\cdot10^{13}$ & $1.12141\cdot10^{12}$ & $0.23$ &  $83$ & $1984$ \\
  MKW4    & - & $4.69503\cdot10^{13}$ & $2.83207\cdot10^{12}$ & $1.71153\cdot10^{11}$ & $5.29855\cdot10^{11}$ & $0.25$ &  $60$ & $ 434$ \\
  RXJ1159 & - & $8.97997\cdot10^{13}$ & $4.33256\cdot10^{12}$ & $7.34414\cdot10^{11}$ & $5.38799\cdot10^{11}$ & $0.29$ &  $64$ & $ 568$ \\
  \hline
 \end{tabular}
 \end{minipage}
\end{table*}

\subsection{Temperature Profiles}
\label{sec:T_prof}

As  stressed in \S~\ref{sec:gas_model}, for the purpose of
this work, we need an accurate qualitative description of the
radial behavior of the gas properties. Standard isothermal or
polytropic models, or even the more complex one proposed in
\cite{Vik06}, do not provide a good description of the data at all
radii and for all clusters in the present sample. We hence describe the
gas temperature profiles using the straightforward X-ray spectral analysis
results, without the introduction of any analytic model.\\
X-ray spectral values have been provided by A. Vikhlinin (private
communication). A detailed description of the relative spectral
analysis can be found in \cite{Vik05}.

\begin{figure}
  \includegraphics[width=84mm]{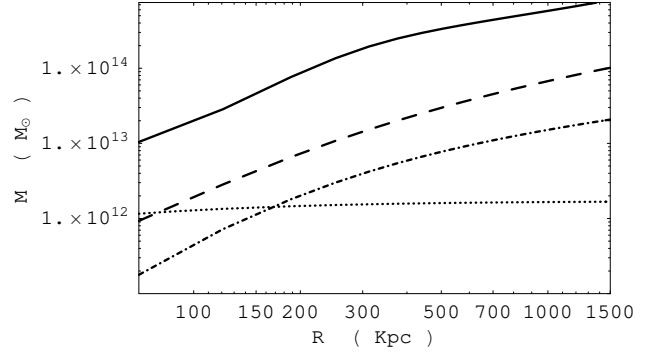}
  \caption{Matter components for A478: total Newtonian dynamical mass (continue
  line); gas mass (dashed line); galactic mass (dotted-dashed line); cD-galaxy mass (dotted line).}
\end{figure}

\subsection{Galaxy Distribution Model}
\label{sec:gal_model}

The galaxy density can be modelled as proposed by \cite{Bah96}.
Even if the galaxy distribution is a \textit{point}-distribution
instead of
 a continuous function, assuming that galaxies are in
equilibrium with gas, we can use a $\beta-$model, $\propto r^{-3}$, for
$r < R_{c}$ from the  cluster center, and a steeper one,
$\propto r^{-2.6}$, for $r > R_{c}$, where $R_{c}$ is the cluster
core radius (its value is taken from Vikhlinin 2006). Its final
expression is:
\begin{equation}\label{gal density bahcall}
\rho_{gal}(r) = \left\{%
\begin{array}{ll}
    \rho_{gal,1} \cdot \left[1+
\left(\frac{r}{R_{c}}\right)^{2} \right]^{-\frac{3}{2}} & \hbox{$r < R_{c}$} \\
    \rho_{gal,2} \cdot \left[1+
\left(\frac{r}{R_{c}}\right)^{2} \right]^{-\frac{2.6}{2}} & \hbox{$r > R_{c}$} \\
\end{array}%
\right.
\end{equation}
where the constants $\rho_{gal,1}$ and $\rho_{gal,2}$ are chosen
in the following way:
\begin{itemize}
 \item \cite{Bah96} provides the central number density of galaxies in
 rich compact clusters for  galaxies located within a
 $1.5$ h$^{-1}$Mpc radius from the cluster center and brighter than $m_3+2^m$
 (where $m_3$ is the magnitude of the third brightest galaxy):
 $n_{gal,0} \sim 10^{3} h^{3}$ galaxies Mpc$^{-3}$. Then we  fix
 $\rho_{gal,1}$ in the range $\sim 10^{34}\div 10^{36}$ kg/kpc$^{3}$.
 For any cluster obeying the condition chosen for the mass
 ratio gal-to-gas, we  assume a typical elliptical and
 cD galaxy mass in the range $10^{12}\div 10^{13} M_{\odot}$.
 \item the constant $\rho_{gal,2}$ has been fixed with the only
 requirement that the galaxy density function has to be continuous at
 $R_{c}$.
\end{itemize}
We have tested the effect of varying galaxy density in the above range
 $\sim 10^{34}\div 10^{36}$ kg/kpc$^{3}$ on the cluster with the lowest mass, namely A262.
 In this case, we would
expect  great variations with respect to  other clusters;
the result is that the contribution due to galaxies and  cD-galaxy
gives a variation $\leq 1\%$ to the final estimate of  fit
parameters. \\
The cD galaxy density has been modelled as described in
\cite{SA06}; they use a Jaffe model of the form:
\begin{equation}\label{jaffe cd galaxy}
\rho_{CDgal} = \frac{\rho_{0,J}}{\left(\frac{r}{r_{c}}\right)^{2}
\left(1+\frac{r}{r_{c}}\right)^{2}}
\end{equation}
where $r_{c}$ is the core radius while the central density is
obtained from ${\displaystyle M_{J} = \frac{4}{3} \pi R_{c}^{3}
\rho_{0,J}}$. The mass of the cD galaxy has been fixed at $1.14
\times 10^{12}$ $M_{\odot}$, with $r_{c} = R_{e}/0.76$, with
$R_{e} = 25$ kpc being the effective radius of the galaxy. The
central galaxy for each cluster in the sample is assumed to have
approximately this stellar mass.

We have assumed that the total galaxy-component mass (galaxies
plus cD galaxy masses) is $\approx 20\div25\%$ of the gas mass: in
\cite{Schindler02}, the mean fraction of gas versus the total mass
(with dark matter) for a cluster is estimated to be $15\div 20\%$,
while the same quantity for galaxies is $3\div 5\%$. This means
that the relative mean mass ratio gal-to-gas in a cluster is
$\approx 20\div 25\%$. We have varied the parameters
$\rho_{gal,1}$, $\rho_{gal,2}$ and $M_{J}$ in their previous
defined ranges to obtain a mass ratio between total galaxy mass
and total gas mass which lies in this range. Resulting galaxy mass
values and ratios ${\displaystyle
\frac{\mathrm{gal}}{\mathrm{gas}}}$, estimated at $r=r_{max}$, are
listed in Table~\ref{tab1}.

In Fig.~(1), we show how each component is spatially distributed.
The CD-galaxy is dominant with respect to the other galaxies only
in the inner region (below $100$ kpc). As already stated in
\S~\ref{sec:gas_model}, cluster innermost regions have been
excluded from our analysis and so the contribution due to the
cD-galaxy is practically negligible in our analysis. The gas is,
as a consequence, clearly the dominant visible component, starting
from innermost regions out to large radii, being galaxy mass only
$20\div 25\%$ of gas mass. A similar behavior is shown by all the
clusters considered in our sample.

\begin{table*}
 \begin{minipage}{152mm}
 \caption{Column 1: Cluster name. Column 2: first derivative coefficient, $a_{1}$, of f(R) series.
          Column3: $1\sigma$ confidence interval for $a_{1}$. Column 4:
          second derivative coefficient, $a_{2}$, of f(R) series. Column 5: $1\sigma$ confidence interval for $a_{2}$.
          Column 6: characteristic length, L, of the modified gravitational potential, derived from $a_{1}$ and $a_{2}$.
          Column 7 : $1\sigma$ confidence interval for $L$.}
 \label{tab2}
 \begin{tabular}{|c|c|c|c|c|c|c|}
  \hline
  name & $a_{1}$ & [$a_{1}-1\sigma$, $a_{1}+1\sigma$] & $a_{2}$ & [$a_{2}-1\sigma$, $a_{2}+1\sigma$] & $L$ & [$L-1\sigma$, $L+1\sigma$]\\
  &  &  & $({\mathrm{kpc}}^{2})$ & $({\mathrm{kpc}}^{2})$ &  (kpc) & (kpc)\\
  \hline
  \hline
  A133    &     $0.085$   &       [$0.078$, $0.091$]    &           $-4.98 \cdot 10^{3}$   &    [$-2.38 \cdot 10^{4}$, $-1.38  \cdot 10^{3}$]   &      $591.78$   &   [$323.34$, $1259.50$]  \\
  A262    &     $0.065$   &       [$0.061$, $0.071$]    &                       $-10.63$   &                              [$-57.65$, $-3.17$]   &       $31.40$   &      [$17.28$, $71.10$]  \\
  A383    &     $0.099$   &       [$0.093$, $0.108$]    &           $-9.01 \cdot 10^{2}$   &     [$-4.10 \cdot 10^{3}$, $-3.14 \cdot 10^{2}$]   &      $234.13$   &    [$142.10$, $478.06$]  \\
  A478    &     $0.117$   &       [$0.114$, $0.122$]    &           $-4.61 \cdot 10^{3}$   &     [$-1.01 \cdot 10^{4}$, $-2.51 \cdot 10^{3}$]   &      $484.83$   &    [$363.29$, $707.73$]  \\
  A907    &     $0.129$   &       [$0.125$, $0.136$]    &           $-5.77 \cdot 10^{3}$   &     [$-1.54 \cdot 10^{4}$, $-2.83 \cdot 10^{3}$]   &      $517.30$   &    [$368.84$, $825.00$]  \\
  A1413   &     $0.115$   &       [$0.110$, $0.119$]    &           $-9.45 \cdot 10^{4}$   &     [$-4.26 \cdot 10^{5}$, $-3.46 \cdot 10^{4}$]   &     $2224.57$   &  [$1365.40$, $4681.21$]  \\
  A1795   &     $0.093$   &       [$0.084$, $0.103$]    &           $-1.54 \cdot 10^{3}$   &     [$-1.01 \cdot 10^{4}$, $-2.49 \cdot 10^{2}$]   &      $315.44$   &    [$133.31$, $769.17$]  \\
  A1991   &     $0.074$   &       [$0.072$, $0.081$]    &                       $-50.69$   &                    [$-3.42 \cdot 10^{2}$, $-13$]   &       $64.00$   &     [$32.63$, $159.40$]  \\
  A2029   &     $0.129$   &       [$0.123$, $0.134$]    &           $-2.10 \cdot 10^{4}$   &     [$-7.95 \cdot 10^{4}$, $-8.44 \cdot 10^{3}$]   &      $988.85$   &   [$637.71$, $1890.07$]  \\
  A2390   &     $0.149$   &       [$0.146$, $0.152$]    &           $-1.40 \cdot 10^{6}$   &     [$-5.71 \cdot 10^{6}$, $-4.46 \cdot 10^{5}$]   &     $7490.80$   & [$4245.74$, $15715.60$]  \\
  MKW4    &     $0.054$   &       [$0.049$, $0.060$]    &                       $-23.63$   &                  [$-1.15 \cdot 10^{2}$, $-8.13$]   &       $51.31$   &     [$30.44$, $110.68$]  \\
  RXJ1159 &     $0.048$   &       [$0.047$, $0.052$]    &                       $-18.33$   &                  [$-1.35 \cdot 10^{2}$, $-4.18$]   &       $47.72$   &     [$22.86$, $125.96$]  \\
  \hline
 \end{tabular}
 \end{minipage}
\end{table*}

\subsection{Uncertainties on mass profiles}
\label{sec:uncertainties}

Uncertainties on the cluster total mass profiles have been
estimated performing Monte-Carlo simulations~\cite{NeuBoh95}. We
proceed to simulate temperature profiles and choose random
radius-temperature values couples for each bin which we have in
our temperature data given by \cite{Vik05}. Random temperature
values have been extracted from a Gaussian distribution centered
on the spectral values, and with a dispersion fixed to its $68\%$
confidence level. For the radius, we choose a random value inside
each bin. We have performed 2000 simulations for each cluster and
perform two cuts on the simulated profile. First, we exclude those
profiles that give an unphysical negative estimate of the mass:
this is possible when our simulated couples of quantities give
rise to too high temperature-gradient. After this cut, we have
$\approx1500$ simulations for any cluster. Then we have ordered
the resulting mass values for increasing radius values. Extreme
mass estimates (outside the $10\div90\%$ range) are excluded from
the obtained distribution, in order to avoid other high mass
gradients which give rise to masses too different from real data.
The resulting limits provide the errors on the total mass.
Uncertainties on the electron-density profiles  have not been
included in the simulations, being them negligible with respect to
those of the gas-temperature profiles.

\subsection{Fitting the mass profiles}
\label{sec:mass_profiles_fit}

In the above sections, we have shown that, with the aid of X-ray
observations, modelling theoretically the galaxy distribution and
using Eq.~(\ref{eq:fit relation}), we obtain an
estimate of the baryonic content of clusters.\\
We have hence performed a best-fit analysis of the theoretical
 Eq.~(\ref{eq:fit relation})
{\setlength\arraycolsep{0.2pt}
\begin{eqnarray}\label{eq:theo_bar}
M_{bar,th}(r) &=& \frac{4 a_{1}}{3} \left[ - \frac{k T(r)}{\mu
m_{p}G} r \left(\frac{d\ln\rho_{gas}(r)}{d\ln r}+\frac{d\ln
T(r)}{d\ln r}\right) \right] + \nonumber \\ &-& \frac{4 a_{1}}{3G}
r^{2} \frac{\mathrm{d}\Phi_{C}}{\mathrm{d}r}(r)
\end{eqnarray}}
versus the observed mass contributions
\begin{equation}\label{eq:obs_bar}
M_{bar,obs}(r) = M_{gas}(r) + M_{gal}(r) + M_{CDgal}(r)
\end{equation}
Since not all the data involved in the above estimate have
measurable errors, we cannot perform an \textit{exact} $\chi$-square
minimization: Actually, we can minimize the quantity:
\begin{equation}
\chi^{2} = \frac{1}{N-n_{p}-1} \cdot \sum_{i=1}^{N}
\frac{(M_{bar,obs}-M_{bar,theo})^{2}}{M_{bar,theo}}
\end{equation}
where $N$ is the number of data and $n_{p} = 2$ the  free
 parameters of the model. We minimize the $\chi$-square using the Markov
Chain Monte Carlo Method (MCMC). For each cluster, we have run
various chains to set the best parameters of the used algorithm,
the Metropolis-Hastings one: starting from an initial parameter
vector $\mathbf{p}$ (in our case ${\mathbf{p}} = (a_{1},a_{2})$),
we generate a new trial point $\mathbf{p'}$ from a tested proposal
density $q(\mathbf{p'},\mathbf{p})$, which represents the
conditional probability to get $\mathbf{p'}$, given $\mathbf{p}$.
This new point is accepted with probability
\[
\alpha({\mathbf{p}}, {\mathbf{p'}}) = min \left\{1,
\frac{L({\mathbf{d}}|{\mathbf{p'}}) P({\mathbf{p'}})
q({\mathbf{p'}},{\mathbf{p}})}{L({\mathbf{d}}|{\mathbf{p}})
P({\mathbf{p}}) q({\mathbf{p}},{\mathbf{p'}})}\ \right\}
\]
where ${\mathbf{d}}$ are the data, $L({\mathbf{d}}|{\mathbf{p'}})
\propto \exp(-\chi^{2}/2)$ is the likelihood function,
$P({\mathbf{p}})$ is the prior on the parameters. In our case, the
prior on the fit parameters is related to Eq.~(\ref{lengths
model}): being $L$ a length, we need to force the ratio $a_1/a_2$
to  be positive. The proposal density is  Gaussian symmetric with
respect of the two vectors $\mathbf{p}$ and $\mathbf{p'}$, namely
$q({\mathbf{p}},{\mathbf{p'}}) \propto \exp(-\Delta p^{2} / 2
\sigma^{2})$, with $\Delta {\mathbf{p}} = {\mathbf{p}} -
{\mathbf{p'}}$; we decide to fix the dispersion $\sigma$ of any
trial distribution of parameters equal to $20\%$ of trial $a_{1}$
and $a_{2}$ at any step. This means that the parameter $\alpha$
reduces to the ratio between the likelihood functions. We have run
one chain of $10^{5}$ points for every cluster; the convergence of
the chains has been tested using the power spectrum analysis from
\cite{Dunkley05}. The key idea of this method is, at the same
time, simple and powerful: if we take the \textit{power spectra}
of the MCMC samples, we will have a great correlation on small
scales but, when the chain reaches convergence, the spectrum
becomes flat (like a white noise spectrum); so that, by checking
the spectrum of just one chain (instead of many parallel chains as
in Gelmann-Rubin test) will be sufficient to assess the reached
convergence. Remanding to \cite{Dunkley05} for a detailed
discussion of all the mathematical steps. Here  we calculate the
discrete power spectrum of the chains:
\begin{equation}\label{eq:disc_power_def}
P_{j} = |a_{N}^{j}|^{2}
\end{equation}
with
\begin{equation}\label{eq:disc_power_coeff}
a_{N}^{j} = \frac{1}{\sqrt{N}}\sum_{n=0}^{N-1} x_{n}
\exp{\left[i\frac{2\pi j}{N}n\right]}
\end{equation}
where $N$ and $x_{n}$ are the length and the element of the sample
from the MCMC, respectively, $j=1,\ldots,\frac{N}{2}-1$. The
wavenumber $k_{j}$ of the spectrum is related to the index $j$ by
the relation $k_{j}=\frac{2\pi j}{N}$. Then we fit it with the
analytical template:
\begin{equation}\label{eq:dunk_template}
P(k) = P_{0} \frac{(k^{*}/k)^{\alpha}}{1+(k^{*}/k)^{\alpha}}
\end{equation}
or in the equivalent logarithmic form:
\begin{equation}
\ln P_{j} = \ln P_{0} + \ln
\left[\frac{(k^{*}/k_{j})^{\alpha}}{1+(k^{*}/k_{j})^{\alpha}}\right]-\gamma
+ r_{j}
\end{equation}
where $\gamma=0.57216$ is the Euler-Mascheroni number and $r_{j}$
are random measurement errors with $<r_{j}> = 0$ and $<r_{i}r_{j}>
= \delta_{ij} \pi^{2}/6$. From the fit, we estimate the two
fundamental parameters, $P_{0}$ and $j^{*}$ (the index
corresponding to $k^{*}$). The first one is the value of the power
spectrum extrapolated for $k \rightarrow 0$ and,  from it, we can
derive the convergence ratio from ${\displaystyle r \approx
\frac{P_{0}}{N}}$; if $r < 0.01$, we can assume that the
convergence is reached. The second parameter is related to the
turning point from a power-law to a flat spectrum. It has to be
$>20$ in order to be sure that the number of points in the sample,
coming from the convergence region, are more than the noise
points. If these two conditions are verified for all the
parameters, then the chain has reached the convergence and the
statistics derived from  MCMC well describes  the underlying
probability distribution (typical results are shown in Figs. \ref{fig:plot_hist_1},
\ref{fig:plot_hist_2}, \ref{fig:plot_hist_3}.
Following \cite{Dunkley05} prescriptions, we perform the
fit over the range $1 \leq j \leq j_{max}$, with $j_{max} \sim 10
j^{*}$, where a first estimation of $j^{*}$ can be obtained from a
fit with $j_{max} = 1000$, and then performing a second iteration
in order to have a better estimation of it. Even if the
convergence is achieved after few thousand steps of the chain, we
have decided to run longer chains of $10^{5}$ points to reduce the
noise from the histograms and avoid under- or over-  estimations
of errors on the parameters. The $i-\sigma$ confidence levels are
easily estimated deriving them from the final sample the
$15.87$-th and $84.13$-th quantiles (which define the $68\%$
confidence interval) for $i=1$, the $2.28$-th and $97.72$-th
quantiles (which define the $95\%$ confidence interval) for $i=2$
and the $0.13$-th and $99.87$-th quantiles (which define the
$99\%$ confidence interval) for $i=3$.

After the description of the method, let us now comment
on the achieved results.

\section{Results}
\label{sec:results}

The numerical results of our fitting analysis are summarized in
Table 2; we give the best fit values of the independent fitting
parameters $a_{1}$ and $a_{2}$, and of the gravitational length
$L$, considered as a function of the previous two quantities. In
 Figs.~\ref{fig:plot_hist_1}-~\ref{fig:plot_hist_3}, we give
the typical results of fitting, with
histograms and power spectrum of samples derived by the MCMC, to
assess the reached convergence (flat spectrum at  large scales).

The goodness and the properties of the fits are shown in
Figs.~\ref{fig:plot_fin_1}-~\ref{fig:plot_fin_12}. The main
property of our results is the presence of a \textit{typical
scale} for each cluster above which our model works really
well (typical relative differences are less than $5\%$), while
for lower scale there is a great difference. It is possible to
see, by a rapid inspection, that this turning-point is located at
a radius $\approx 150$ kpc. Except for extremely rich clusters, it
is  clear that this value is independent of the cluster, being
similar for all clusters in our sample.

There are two main independent explanations that could justify
this trend: limits due to  a break in the state of hydrostatic
equilibrium or limits in the series expansion of the
$f(R)$-models.

If the hypothesis of hydrostatic equilibrium is not correct, then
we are in a regime where the fundamental relations
Eqs.~(\ref{Boltzmann equation})-~(\ref{eq:Boltzmann potential}),
are not working. As discussed in \cite{Vik05},  the central (70
kpc) region of  most clusters is strongly affected by radiative
cooling and thus  its physical properties cannot directly be
related to the depth of the cluster potential well. This means
that, in this region, the gas is not in hydrostatic equilibrium
but in a  multi-phase state.
In this case, the gas temperature cannot be used as a good
standard tracer.

We have also to consider another limit of our modelling: the
requirement that the $f(R)$-function is Taylor expandable. The
corrected gravitational potential which we have considered is
derived in the weak field limit, which means
\begin{equation}
R - R_0 << \frac{a_{1}}{a_{2}}
\end{equation}
where $R_0$ is the background value of the curvature. If this
condition is not satisfied, the approach does not work (see
\cite{noi-prd} for a detailed discussion of this point).
Considering that $a_{1}/a_{2}$ has the dimension of $length^{-2}$
this condition  defines the \textit{length scale} where our series
approximation can work. In other words, this indicates the limit
in which the model can be compared with  data.

For the considered sample, the fit of the parameters $a_{1}$ and
$a_{2}$, spans the length range $\{19; 200\}$ kpc (except for the
richest clusters). It is evident that every galaxy cluster has a
\textit{proper} gravitational length scale. It is worth noticing
that a similar situation, but at completely different scales, has
been found out for low surface brightness galaxies modelled by
$f(R)$-gravity \cite{CapCardTro07}.

Considering the  data at our disposal and the analysis which we
have performed, it is not possible to quantify exactly the
quantitative amount of  these two different phenomena (i.e. the
radiative cooling and the validity of the weak field limit).
However, they are not mutually exclusive but should be considered
in details in view of a more refined modelling \footnote{Other
secondary phenomena as cooling flows, merger and asymmetric shapes
have to be considered in view of a detailed modelling of clusters.
However, in this work, we are only interested  to show that
extended gravity could be a valid alternative to dark matter in
order to explain the  cluster dynamics. }.

Similar issues are present also in \cite{Brown06}: they use the
the Metric - Skew - Tensor - Gravity (MSTG) as a generalization of
the Einstein General Relativity and derive the gas mass profile of
a sample of clusters with gas being the only baryonic component of
the clusters. They consider some clusters  included in our sample
 (in particular, A133, A262, A478, A1413, A1795, A2029, MKW4) and they find the
same different trend for $r \leq 200$ kpc, even if with a
different behavior with respect to us: our model gives lower
values than X-ray gas mass data while their model gives higher
values with respect to X-ray gas mass data.  This stresses the
need for a more accurate modelling of the gravitational potential.

\begin{figure}
\centering
  \includegraphics[width=84mm]{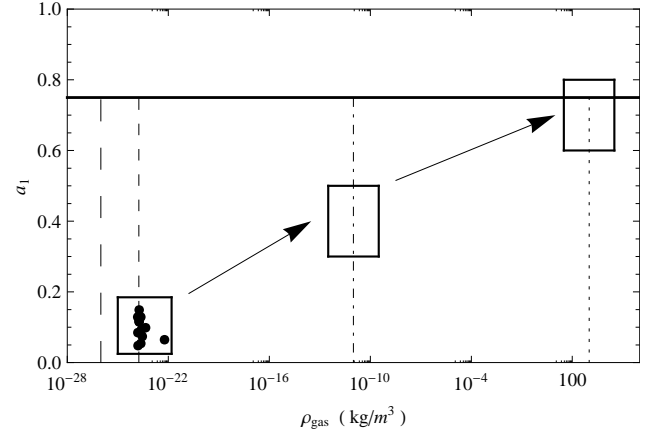}
  \caption{Density vs $a_{1}$: predictions on the behavior of
  $a_{1}$. The horizontal black bold line indicates the
  Newtonian-limit, $a_{1} \rightarrow 3/4$ which we expect to be realized on
  scales comparable with Solar System. Vertical lines indicate typical
  approximated values of matter density (without dark matter) for different
  gravitational structures: universe (large dashed) with critical
  density $\rho_{crit} \approx 10^{-26}$ $\mathrm{kg/m^{3}}$; galaxy clusters (short
  dashed) with $\rho_{cl} \approx 10^{-23}$ $\mathrm{kg/m^{3}}$; galaxies (dot-dashed)
  with $\rho_{gal} \approx 10^{-11}$ $\mathrm{kg/m^{3}}$; sun (dotted) with
  $\rho_{sun} \approx 10^{3}$ $\mathrm{kg/m^{3}}$. Arrows and boxes show the predicted
  trend for $a_{1}$.}
\label{fig:plots_finali1}
\end{figure}

However, our goal is to show that potential (\ref{gravpot1}) is
suitable to fit the mass profile of galaxy clusters and that it
comes from a self-consistent theory.

In general, it can be shown  that the weak field limit of
extended theories of gravity has Yukawa-like corrections \cite{b10,kenmoku}.
Specifically, given theory of gravity of order $(2n+2)$,
the Yukawa corrections to the Newtonian potential are $n$ \cite{Qua91}.
This means that if the effective Lagrangian of the theory is
\begin{equation}
{\cal L}=f(R,\Box R,..\Box^k R,..\Box^n R)\sqrt{-g}
\end{equation}
we have
  \begin{equation} \label{yukawa} \phi(r)=-\frac{G
M}{r}\left[1+\sum_{k=1}^{n}\alpha_k e^{-r/L_k}\right]\,.
\end{equation}
Standard General Relativity, where Yukawa corrections are not present,
is recovered for $n=0$ (second order theory) while the
 $f(R)$-gravity is obtained for $n=1$ (fourth-order theory). Any $\Box$ operator introduces two
further derivation orders in the field equations.
This kind of Lagrangian comes out when  quantum field theory is formulated on  curved spacetime \cite{birrell}. In the series (\ref{yukawa}),
$G$ is the value of the gravitational
constant considered at infinity, $L_k$ is the interaction length
of the $k$-th component of the non-Newtonian corrections. The
amplitude $\alpha_k$ of each component is normalized to the
standard Newtonian term; the sign of $\alpha_k$ tells us if the
corrections are attractive or repulsive (see \cite{will} for
details). Moreover, the variation of the gravitational coupling is
involved. In our case, we are taking into account only the first
term of the series. It is the
the leading term. Let us rewrite
(\ref{gravpot1}) as

\begin{equation}
\label{yukawa1} \phi(r)=-\frac{G M}{r}\left[1+\alpha_1
e^{-r/L_1}\right]\,. \end{equation} The effect of non-Newtonian
term can be parameterized by $\{\alpha_1,\,L_1\}$ which could be a useful
parameterisation which respect to our previous
 $\{a_1,\,a_2\}$ or $\{G_{eff},\,L\}$ with
$G_{eff}=3G/(4a_1)$. For large distances, where $r\gg L_1$, the
exponential term vanishes and the gravitational coupling is $G$.
If $r\ll L_1$, the exponential becomes 1 and, by differentiating
Eq.(\ref{yukawa1}) and comparing with the gravitational force
measured in laboratory, we get

\begin{equation} \label{yukawa2}
G_{lab}=G\left[1+\alpha_1\left(1+\frac{r}{L_1}\right)e^{-r/L_1}\right]
\simeq G(1+\alpha_1)\,, \end{equation} where $G_{lab}=6.67\times
10^{-8}$ g$^{-1}$cm$^3$s$^{-2}$ is the usual Newton constant
measured by Cavendish-like experiments. Of course, $G$ and
$G_{lab}$ coincide in the standard Newtonian gravity. It is worth noticing
that, asymptotically, the inverse square law holds but the
measured coupling constant differs by a factor $(1+\alpha_1)$. In
general, any  correction introduces a characteristic length that
acts at a certain scale for the self-gravitating systems as in the
case of galaxy cluster which we are examining here. The range of
$L_k$ of the $k$th-component of non-Newtonian force can be
identified with the mass $m_k$ of a pseudo-particle whose
effective Compton's length can be defined as

\begin{equation} \label{27} L_k=\frac{\hbar}{m_k c}\,. \end{equation} The
interpretation of this fact is that, in the weak energy limit,
fundamental theories which attempt to unify gravity with the other
forces introduce, in addition to the massless graviton, particles
{\it with mass} which also  carry the gravitational interaction
\cite{gibbons}. See, in particular, \cite{PPNfelix} for
$f(R)$-gravity. These masses are related to effective length scales
which can be parameterized as

\begin{equation}\label{28} L_k=2\times 10^{-5}\left(\frac{1\,
\mbox{eV}}{m_k}\right)\mbox{cm}\,. \end{equation} There have been
several attempts to experimentally constrain $L_k$ and $\alpha_k$ (and then
$m_k$) by experiments on scales in the range $1 \,\mbox{cm}<r<
1000\,\mbox{km}$, using different techniques
\cite{fischbach,speake,eckhardt}. In this case, the expected masses of
particles which should carry the additional gravitational force
are in the range $10^{-13} \mbox{eV}<m_k< 10^{-5}\, \mbox{eV}$.
The general outcome of these experiments, even retaining only the
term $k=1$, is that {\it geophysical window} between the laboratory
and the astronomical scales has to be taken into account. In fact,
the range

\begin{equation} |\alpha_1|\sim 10^{-2}\,,\qquad L_1\sim 10^2\div
10^3\,\,\mbox{m}\,,\end{equation} is not excluded at all in this
window. An interesting suggestion has been given by Fujii
\cite{fujii1}, which proposed that the exponential deviation from
the Newtonian standard potential  could arise from the microscopic
interaction which couples  the nuclear isospin and the baryon number.

The astrophysical counterparts of these non-Newtonian corrections
seemed ruled out till some years ago due to the fact that
experimental tests of General Relativity seemed to predict  the
Newtonian potential in the weak energy limit, ''inside" the Solar
System. However, as it has been shown, several alternative
theories seem to evade the Solar System constraints (see
\cite{PPNfelix} and the reference therein for recent results) and, furthermore, indications
of an anomalous, long--range acceleration revealed from the data
analysis of Pioneer 10/11, Galileo, and Ulysses spacecrafts (which
are now almost outside the Solar System) makes these Yukawa--like
corrections come again into play \cite{anderson}.
Besides, it is
possible to reproduce   phenomenologically the flat rotation
curves of spiral galaxies  considering the values

\begin{equation}\label{sand} \alpha_1=-0.92\,,\qquad L_1\sim
40\,\,\mbox{kpc}\,.\end{equation} The main hypothesis of this
approach is that the additional gravitational interaction is
carried by some ultra-soft boson whose range of mass is $m_1\sim
10^{-27}\div 10^{-28}$eV. The action of this boson becomes
efficient at galactic scales without the request of enormous
amounts of dark matter to stabilize the systems \cite{sanders}.

Furthermore, it is possible to  use a combination of two
exponential correction terms and give a detailed explanation of
the kinematics of galaxies and galaxy clusters, again without dark
matter model \cite{eckhardt}.

It is worthwhile to note that both the spacecrafts measurements
and galactic rotation curves indications come from ''outside" the
usual Solar System boundaries used up to now to test General
Relativity. However, the above results {\it do not come} from any
fundamental theory  to explain the outcome of Yukawa
corrections. In their contexts, these terms are phenomenological.

Another important remark in this direction deserves the fact that
some authors \cite{mcgaugh} interpret also the  experiments on
cosmic microwave background like the experiment BOOMERANG and WMAP
\cite{Boom,WMAP} in the framework of {\it modified
Newtonian dynamics} again without invoking any dark matter model.

All these facts point towards the line of thinking that also
corrections to the standard gravity have to be seriously taken
into account beside dark matter searches.

In our case, the
parameters $a_{1,2}$, which determine the gravitational correction
and the gravitational coupling, come out "directly" from a field theory with
the only requirement that the effective action of gravity
could be  more general than the Hilbert-Einstein theory
$f(R)=R$. This main hypothesis comes from fundamental physics
motivations due to the fact that any unification scheme
or quantum field theory on curved space have
to take into account higher order terms in curvature invariants \cite{birrell}.
Besides, several recent results point out that
such corrections have a main role also at astrophysical and cosmological scales.
For a detailed discussion, see \cite{odi,capfra,faraoni}.

With this philosophy in mind,  we have plotted the trend of $a_{1}$ as a
function of the density in Fig.\ref{fig:plots_finali1}.
As one can see, its values are strongly
constrained in a narrow region of the parameter space, so that
$a_{1}$ can be considered a "tracer" for the size of gravitational
structures. The value of $a_{1}$  range between  $\{0.8 \div 0.12
\}$ for larger clusters and $\{0.4\div 0.6 \}$ for  poorer
structures (i.e. galaxy groups like MKW4 and RXJ1159). We expect a
particular trend when applying the model to different
gravitational structures. In Fig.~\ref{fig:plots_finali1},  we
give characteristic values of density which range from the biggest
 structure, the observed Universe (large dashed vertical line), to the
smallest one, the Sun (vertical dotted line), through intermediate
steps like clusters (vertical short dashed line) and galaxies
(vertical dot-dashed line). The bold black horizontal line
represents the Newtonian limit  $a_{1}=3/4$ and the  boxes
indicate the possible values of $a_{1}$ that we obtain by applying
our theoretical model to different structures.

Similar considerations hold also for the characteristic
gravitational length $L$ directly related to both $a_{1}$ and
$a_{2}$. The parameter $a_{2}$ shows a very large range of
variation $\{-10^{6} \div -10 \}$ with respect to the density (and
the mass)  of the clusters. The value of $L$ changes with the
sizes of gravitational structure (see Fig.~\ref{fig:PropPhysVSL}),
so it can be considered, beside the Schwarzschild radius, a sort of additional gravitational
radius. Particular care must be taken when considering Abell~2390, which shows large cavities in the
X-ray surface brightness distribution, and whose central region, highly asymmetric, is not expected to be in hydrostatic equilibrium.
All results at small and medium radii for this cluster could hence be strongly biased by these
effects~\cite{Vik06}; the same will hold for the resulting exceptionally high value of $L$.
 Fig.~\ref{fig:PropPhysVSL} shows how observational properties of the cluster, which well characterize its
gravitational potential (such as the average temperature and the total cluster mass within
r$_{500}$, plotted in the left and right panel, respectively), well correlate with the characteristic gravitational length $L$.

For clusters, we can define a gas-density-weighted  and a
gas-mass-weighted mean, both depending on the series parameters
$a_{1,2}$. We have:
\begin{eqnarray}
<L>_{\rho} &=& 318 \; \mathrm{kpc} \qquad <a_{2}>_{\rho} = -3.40
\cdot 10^{4} \nonumber \\
<L>_{M} &=& 2738 \; \mathrm{kpc} \qquad <a_{2}>_{M} = -4.15 \cdot
10^{5}
\end{eqnarray}
It is straightforward to note the correlation with the sizes of
the cluster cD-dominated-central region and the "gravitational"
interaction length of the whole cluster. In other words, the
parameters $a_{1,2}$, directly related to the first and second
derivative of a given analytic $f(R)$-model determine the
characteristic sizes of the self gravitating structures.

\section{Discussion and Conclusions}
\label{sec:discussion}

In this paper we have investigated the possibility that the high
observational   mass-to-light ratio of galaxy clusters  could be
addressed by $f(R)$- gravity  without assuming huge amounts of
dark matter. We point out  that this proposal comes out from the
fact that, up to now, no definitive candidate for dark-matter has
been observed at fundamental level and then  alternative solutions
to the problem should be viable. Furthermore, several results in
$f(R)$-gravity seem to confirm that valid alternatives to
$\Lambda$CDM can be achieved in cosmology. Besides, as discussed
in the  Introduction, the rotation curves of spiral galaxies can be
explained in the weak field limit of $f(R)$-gravity.  Results of
our analysis go in this direction.

We have chosen a sample of  relaxed galaxy clusters  for which
accurate spectroscopic temperature measurements and gas mass profiles are available.
For the sake of simplicity, and
considered the sample at our disposal, every cluster has been
modelled as a self bound gravitational system with spherical
symmetry and in hydrostatic equilibrium.  The mass distribution
has been described by a corrected  gravitational  potential
obtained from a generic analytic $f(R)$-theory. In fact, as soon
as $f(R)\neq R$, Yukawa-like exponential corrections emerge in the
weak field limit while the standard Newtonian potential is
recovered only for $f(R)=R$, the Hilbert-Einstein theory.

Our goal has been to analyze if the dark-matter content of
clusters can be addressed by these  correction potential terms. As
discussed in detail in the previous section and how it is possible
to see by a rapid inspection of
Figs.~\ref{fig:plot_fin_1}-~\ref{fig:plot_fin_12}, all the
clusters of the sample are consistent with the proposed model at
$1\sigma$ confidence level.  This shows, at least
\textit{qualitatively},  that the high mass-to-light ratio of
clusters can be explained by using a modified gravitational
potential. The good agreement is achieved on distance scales
starting from $150$ kpc up to $1000$ kpc. The differences observed
at smaller scales can be ascribed  to non-gravitational phenomena,
such as cooling flows, or to the fact that the gas mass is not a
good tracer at this scales. The remarkable result is that we have
obtained a consistent agreement with data only using the corrected
gravitational potential in a large range of radii. In order to put
in evidence this trend, we have plotted the baryonic mass vs radii
considering, for each cluster,  the scale where the trend is
clearly evident.

In our knowledge, the fact that $f(R)$-gravity could work at these
scales  has been only supposed but never achieved by a direct
fitting with data (see \cite{lobo} for a review). Starting from
the series coefficients $a_{1}$ and $a_{2}$, it is possible to
state that, at cluster scales, two characteristic sizes emerge
from the weak field limit of the theory.  However, at smaller
scales, e.g. Solar System scales, standard Newtonian gravity has
to be dominant in agreement with observations.

In conclusion, if our considerations are right, gravitational
interaction depends on the scale and the {\it infrared limit} is
led by the series coefficient of the considered effective
gravitational Lagrangian. Roughly speaking, we expect that
starting from cluster scale to galaxy scale, and then down to
smaller  scales as Solar System or Earth, the terms of the series
lead the clustering of self-gravitating systems beside other
non-gravitational phenomena. In our case, the Newtonian limit is
recovered for $a_{1} \rightarrow 3/4$ and $L(a_{1},a_{2})\gg r$
at small scales and for $L(a_{1},a_{2})\ll r$ at large scales. In the first case, the gravitational coupling has to be redifined, in the second $G_{\infty}\simeq G$.
In these limits, the linear Ricci term is dominant in the
gravitational Lagrangian and the Newtonian gravity is restored
\cite{Qua91}. Reversing the argument, this could be the starting
point to achieve a theory  capable of explaining the strong
segregation in masses and sizes of gravitationally-bound systems.

\section{Acknowledgments}

We warmly thank V.F. Cardone  for priceless help in computational
work. We acknowledge A. Stabile and A. Troisi for comments,
discussions and suggestions on the topic and M. Paolillo  for
hints and suggestions in cluster modelling. We are indebted with
A. Vikhlinin which kindly gave us  data on cluster temperature
profiles.

\begin{figure*}
\centering
  \includegraphics[width=85mm]{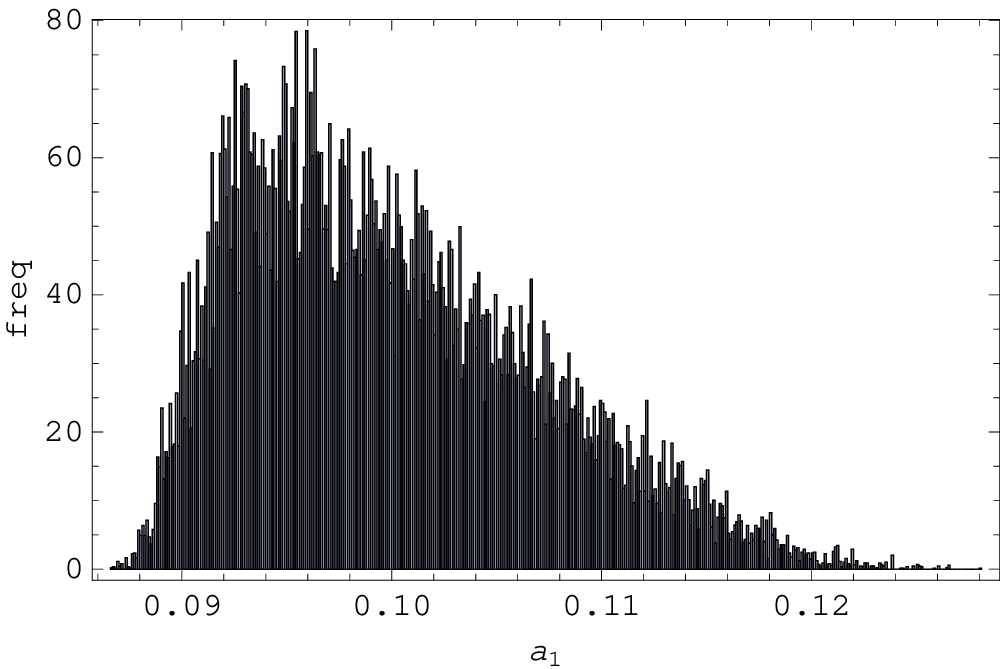}
  \includegraphics[width=85mm]{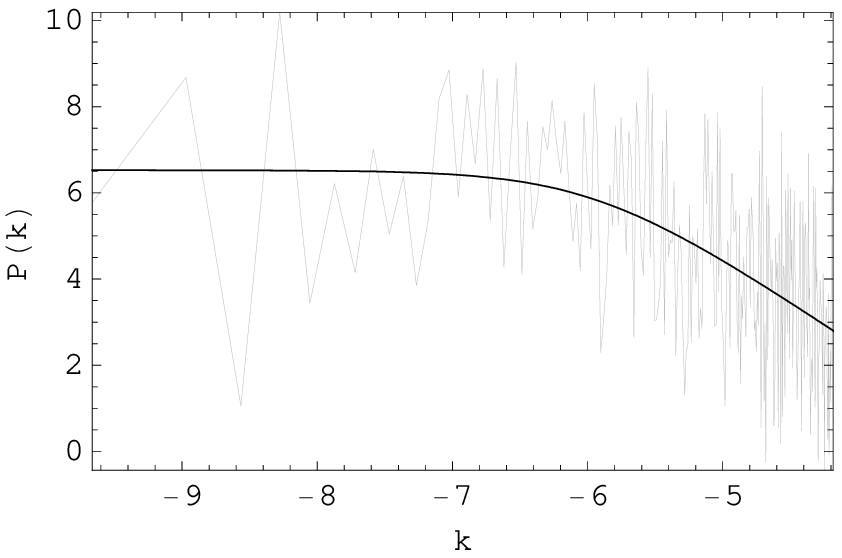}
  \caption{Left panel: histogram of the sample points for parameter $a_{1}$ in Abell 383 coming out the MCMC implementation
  used to estimate best fit values and errors for our fitting procedure as described in
  \S~\ref{sec:mass_profiles_fit}. Binning (horizontal axis) and relative
  frequencies (vertical axis) are given by automatic procedure
  from Mathematica6.0. Right panel: power spectrum test on sample chain for
  parameter $a_{1}$ using the method described in \S~\ref{sec:mass_profiles_fit}. Black line
  is the logarithm of the analytical template Eq.~(\ref{eq:dunk_template}) for power spectrum; gray line is the discrete power spectrum
  obtained using Eq.~(\ref{eq:disc_power_def}) -~(\ref{eq:disc_power_coeff}).}
\label{fig:plot_hist_1}
\end{figure*}

\begin{figure*}
\centering
  \includegraphics[width=85mm]{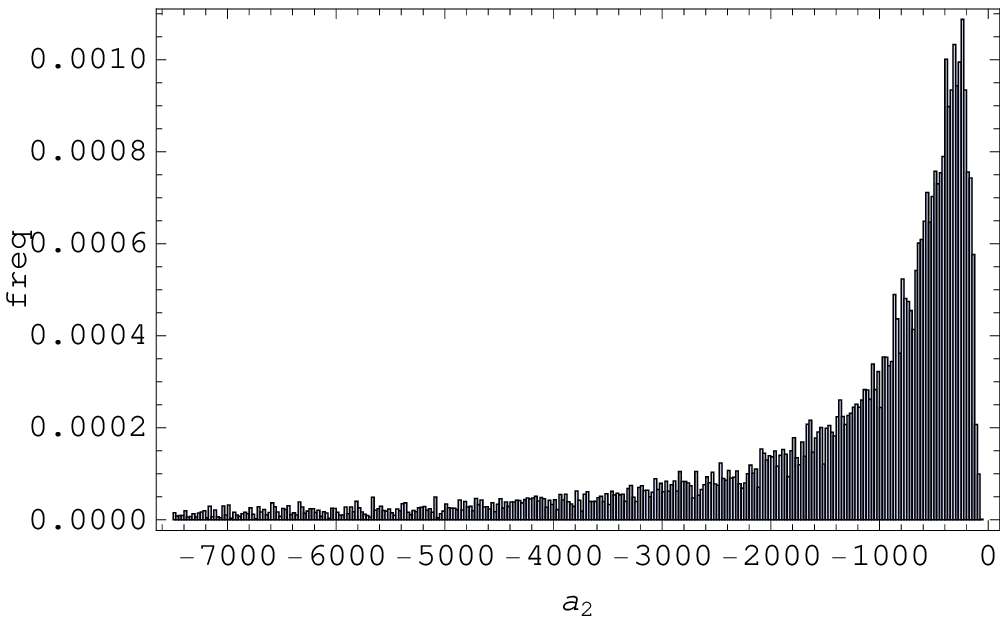}
  \includegraphics[width=85mm]{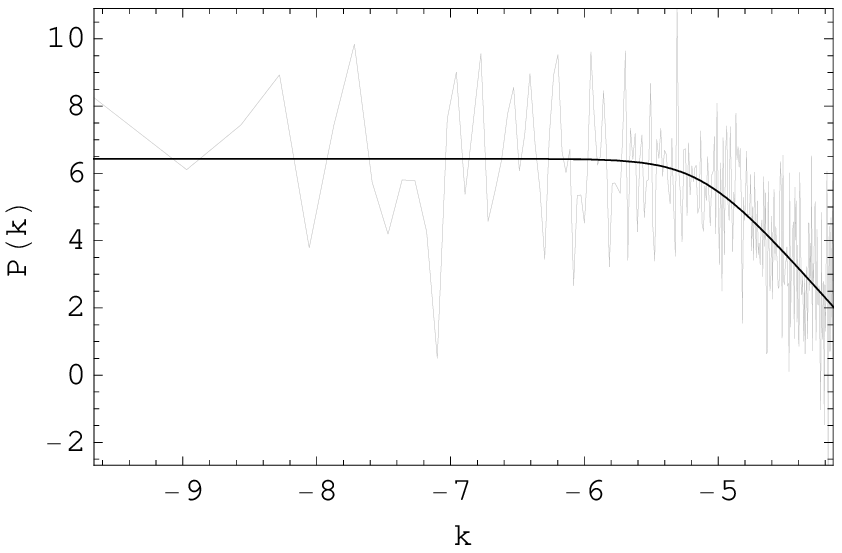}
  \caption{Abell 383: histogram (left) and power spectrum test (right) on sample chain for parameter $a_{2}$.}
\label{fig:plot_hist_2}
\end{figure*}

\begin{figure*}
\centering
  \includegraphics[width=85mm]{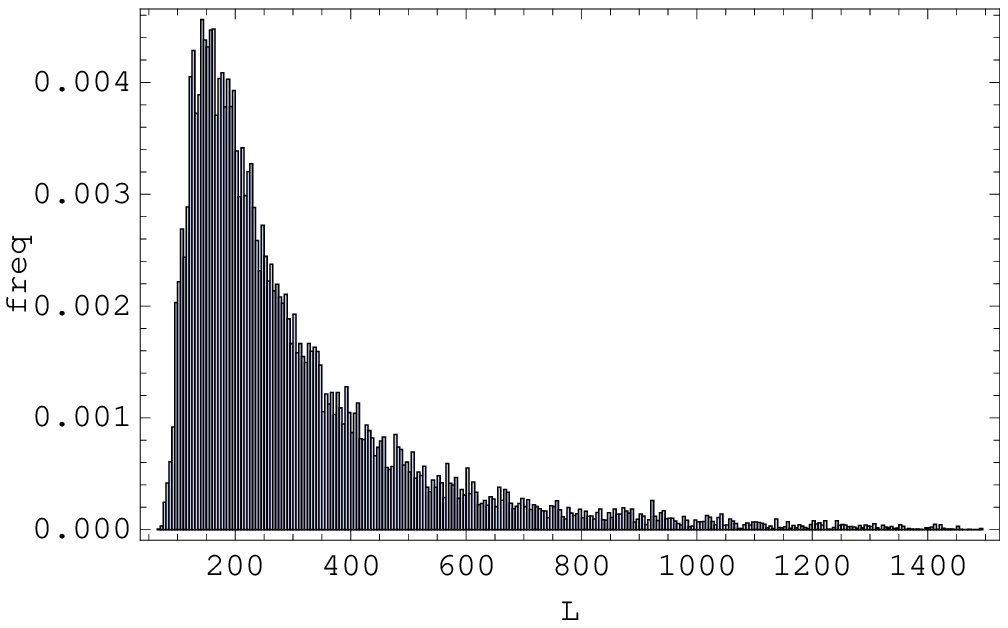}
  \includegraphics[width=85mm]{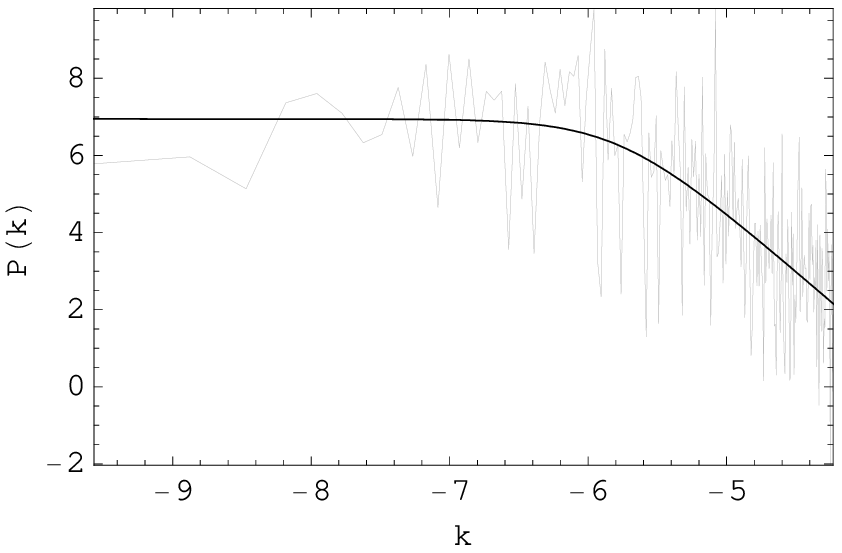}
  \caption{Abell 383: histogram (left) and power spectrum test (right) on sample chain for parameter $L$.}
\label{fig:plot_hist_3}
\end{figure*}

\begin{figure*}
\centering
  \includegraphics[width=85mm]{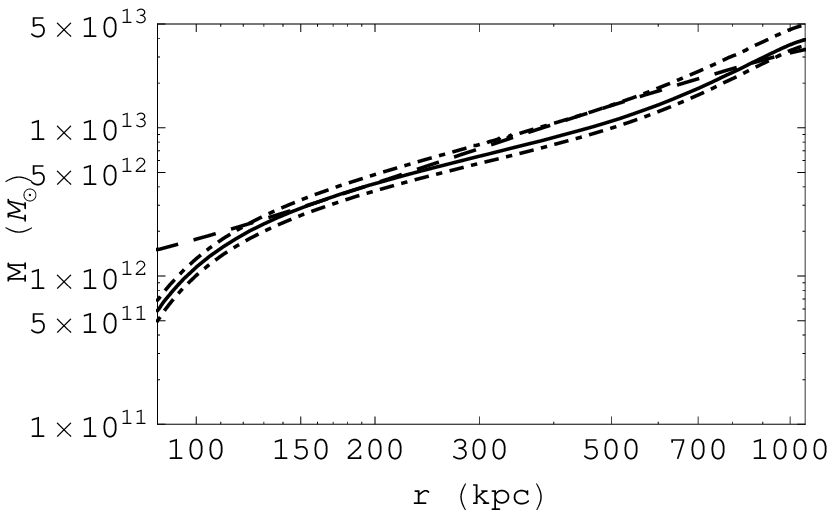}
  \includegraphics[width=85mm]{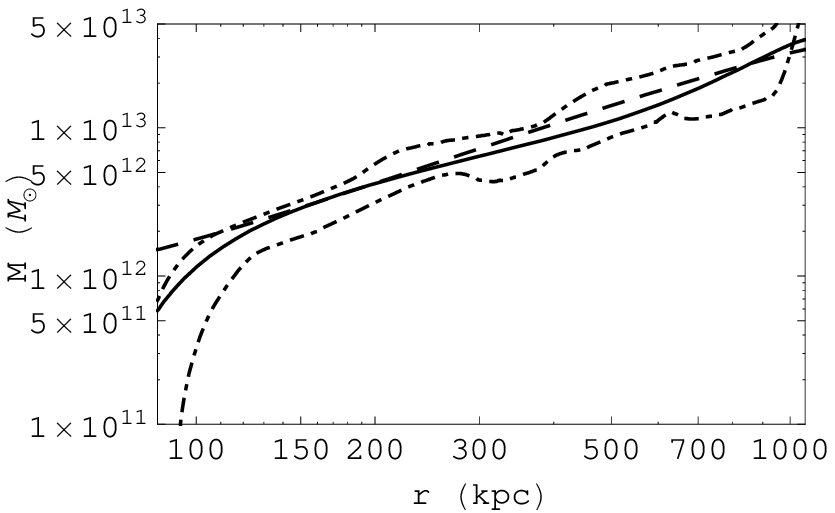}
  \caption{Baryonic mass vs radii for Abell A133. Dashed line is the experimental-observed estimation Eq.~(\ref{eq:obs_bar}) of
  baryonic matter component (i.e. gas, galaxies and cD-galaxy); solid line is the theoretical estimation Eq.~(\ref{eq:theo_bar})
  for baryonic matter component. Dotted lines are the 1-$\sigma$ confidence levels given by errors on fitting
  parameters in the left panel; and from fitting parameter plus statistical errors on mass profiles as
  discussed in \S~\ref{sec:uncertainties} in the right panel.}
\label{fig:plot_fin_1}
\end{figure*}

\begin{figure*}
\centering
  \includegraphics[width=85mm]{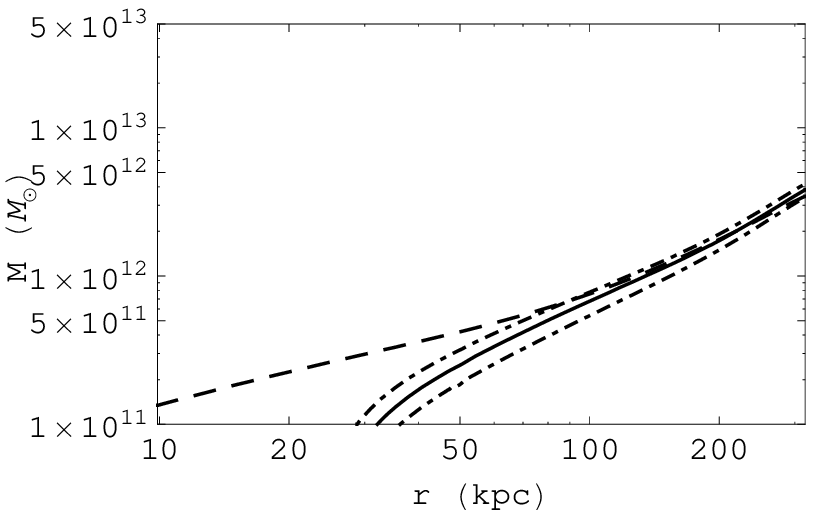}
  \includegraphics[width=85mm]{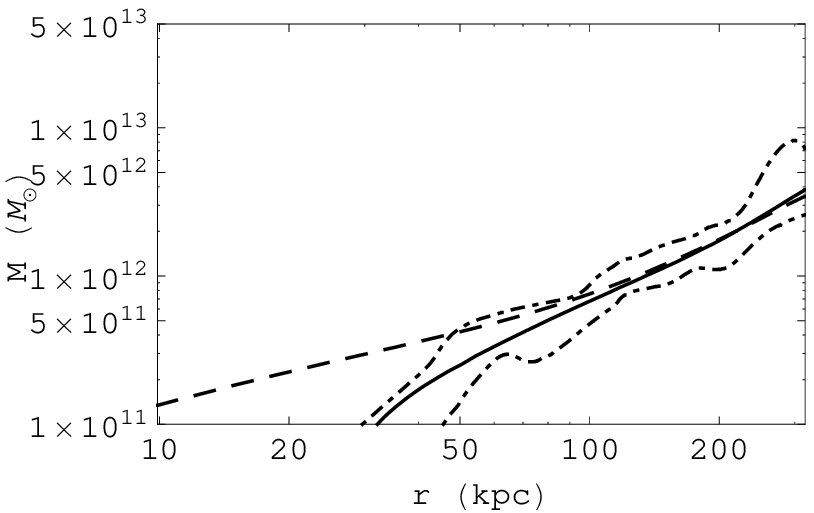}
  \caption{Same of Fig.~\ref{fig:plot_fin_1} but for cluster Abell 262.}
\label{fig:plot_fin_2}
\end{figure*}

\begin{figure*}
\centering
  \includegraphics[width=85mm]{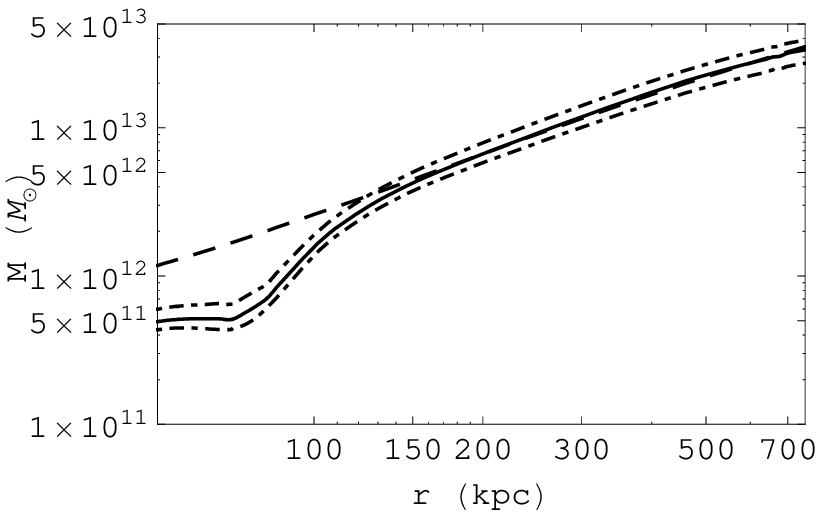}
  \includegraphics[width=85mm]{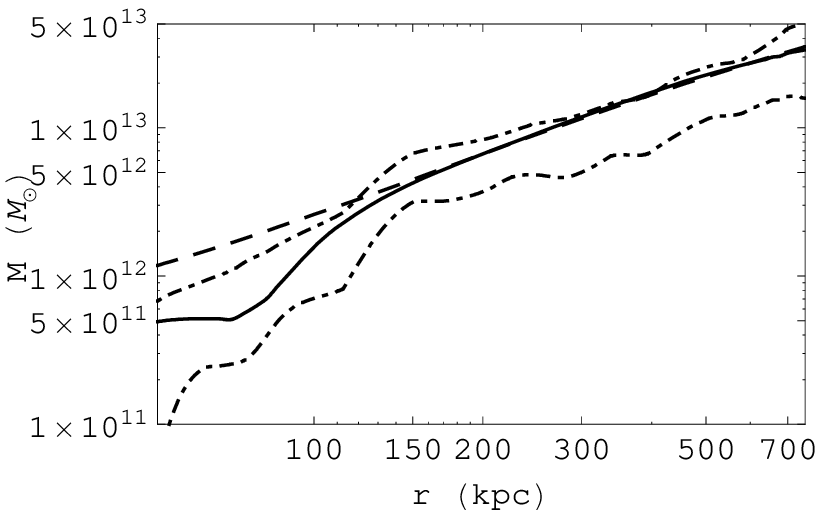}
  \caption{Same of Fig.~\ref{fig:plot_fin_1} but for cluster Abell 383.}
\label{fig:plot_fin_3}
\end{figure*}

\begin{figure*}
\centering
  \includegraphics[width=85mm]{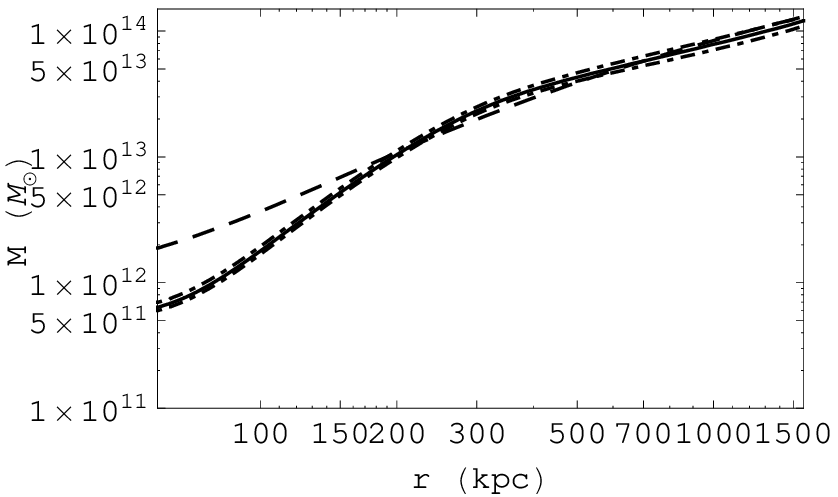}
  \includegraphics[width=85mm]{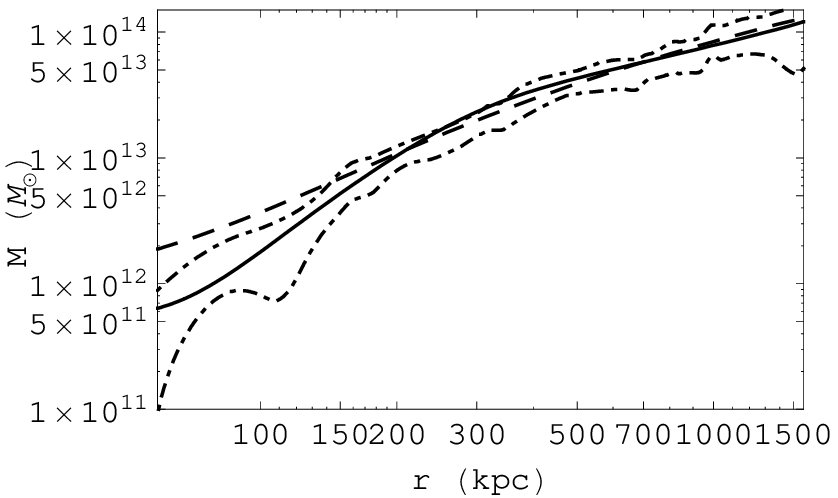}
  \caption{Same of Fig.~\ref{fig:plot_fin_1} but for cluster Abell 478.}
\label{fig:plot_fin_4}
\end{figure*}

\begin{figure*}
\centering
  \includegraphics[width=85mm]{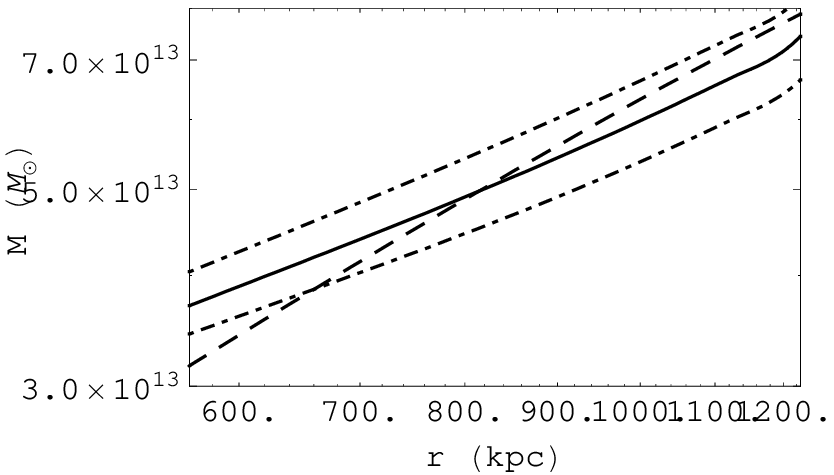}
  \includegraphics[width=85mm]{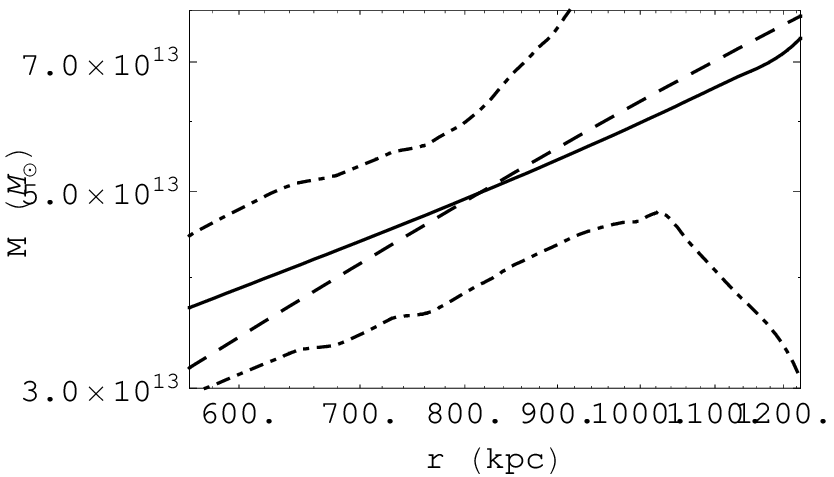}
  \caption{Same of Fig.~\ref{fig:plot_fin_1} but for cluster Abell 907.}
\label{fig:plot_fin_5}
\end{figure*}

\begin{figure*}
\centering
  \includegraphics[width=85mm]{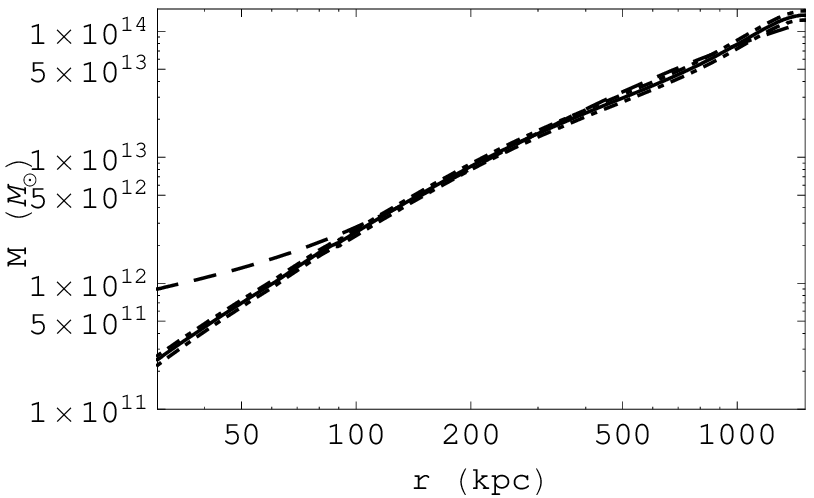}
  \includegraphics[width=85mm]{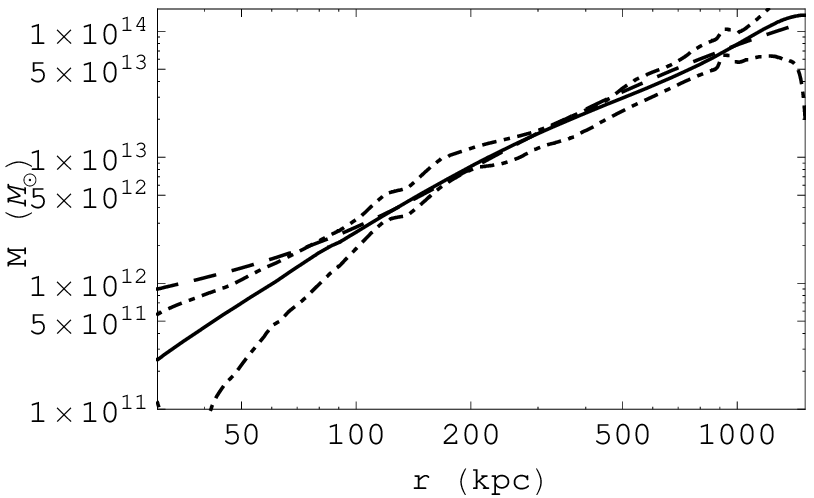}
  \caption{Same of Fig.~\ref{fig:plot_fin_1} but for cluster Abell 1413.}
\label{fig:plot_fin_6}
\end{figure*}

\begin{figure*}
\centering
  \includegraphics[width=85mm]{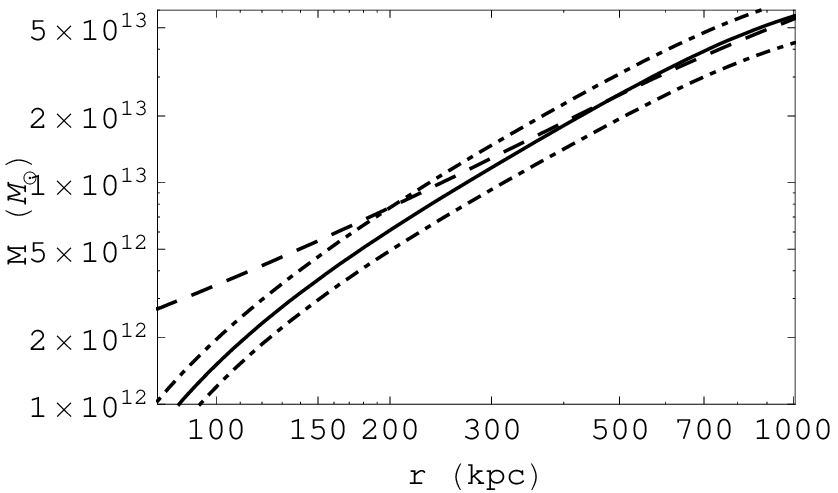}
  \includegraphics[width=85mm]{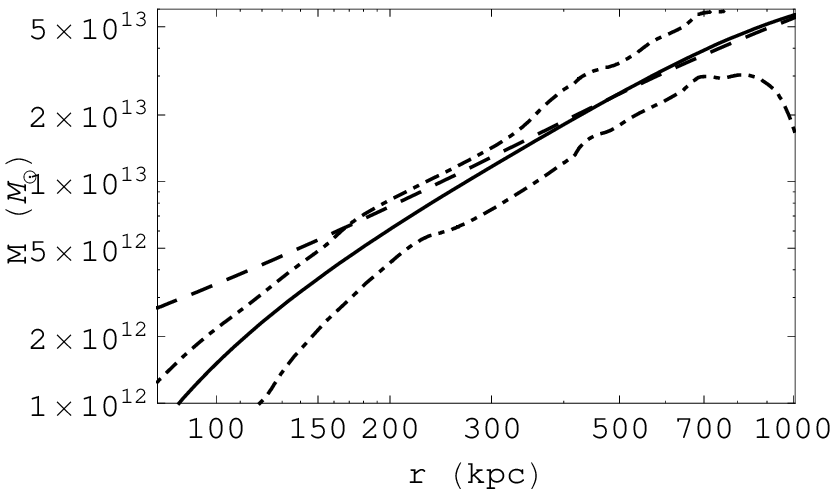}
  \caption{Same of Fig.~\ref{fig:plot_fin_1} but for cluster Abell 1795.}
\label{fig:plot_fin_7}
\end{figure*}

\begin{figure*}
\centering
  \includegraphics[width=85mm]{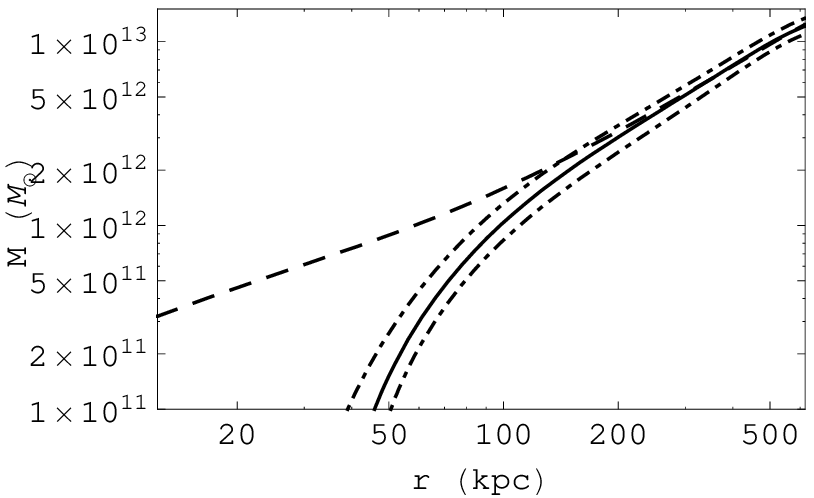}
  \includegraphics[width=85mm]{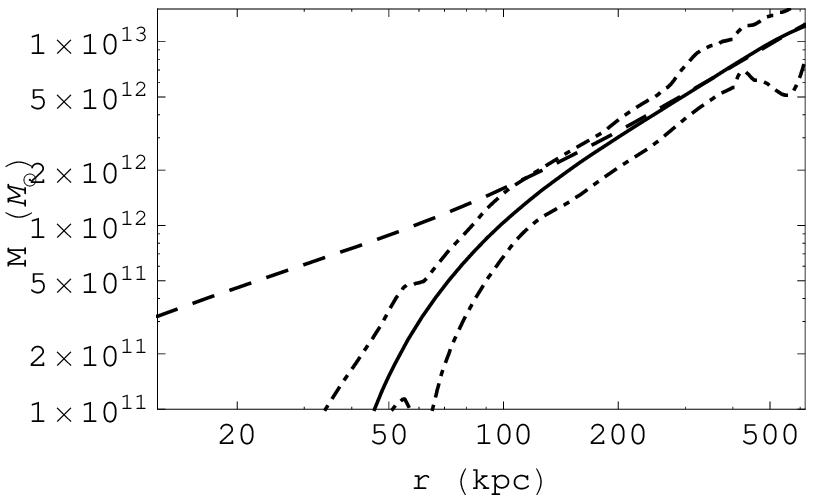}
  \caption{Same of Fig.~\ref{fig:plot_fin_1} but for cluster Abell 1991.}
\label{fig:plot_fin_8}
\end{figure*}

\begin{figure*}
\centering
  \includegraphics[width=85mm]{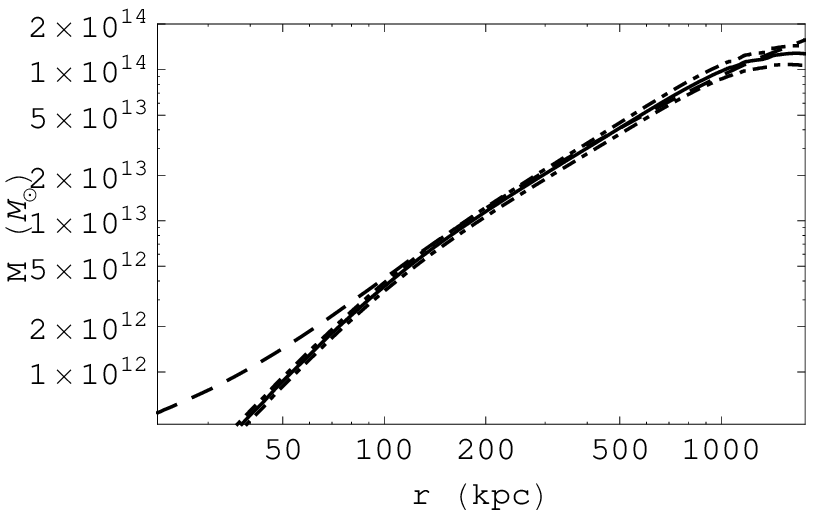}
  \includegraphics[width=85mm]{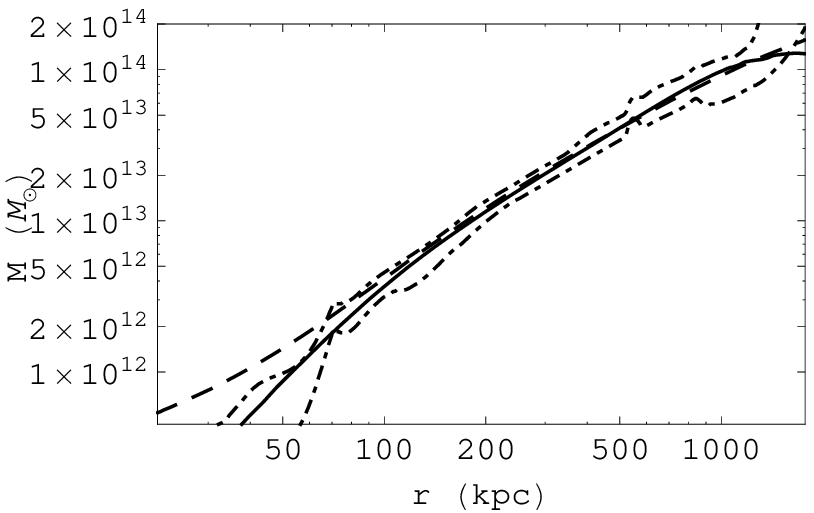}
  \caption{Same of Fig.~\ref{fig:plot_fin_1} but for cluster Abell 2029.}
\label{fig:plot_fin_9}
\end{figure*}

\begin{figure*}
\centering
  \includegraphics[width=85mm]{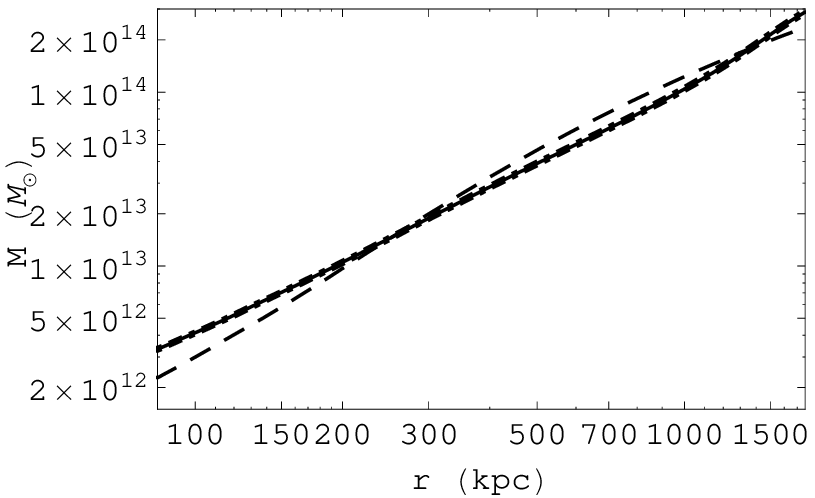}
  \includegraphics[width=85mm]{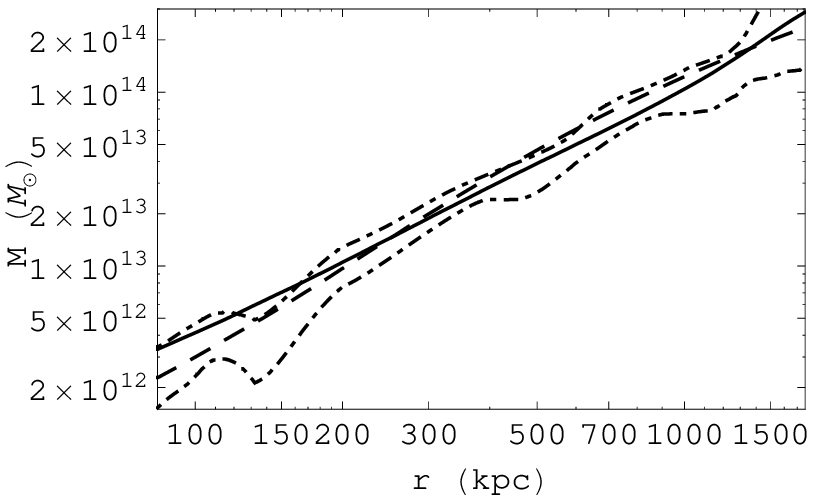}
  \caption{Same of Fig.~\ref{fig:plot_fin_1} but for cluster Abell 2390.}
\label{fig:plot_fin_10}
\end{figure*}

\begin{figure*}
\centering
  \includegraphics[width=85mm]{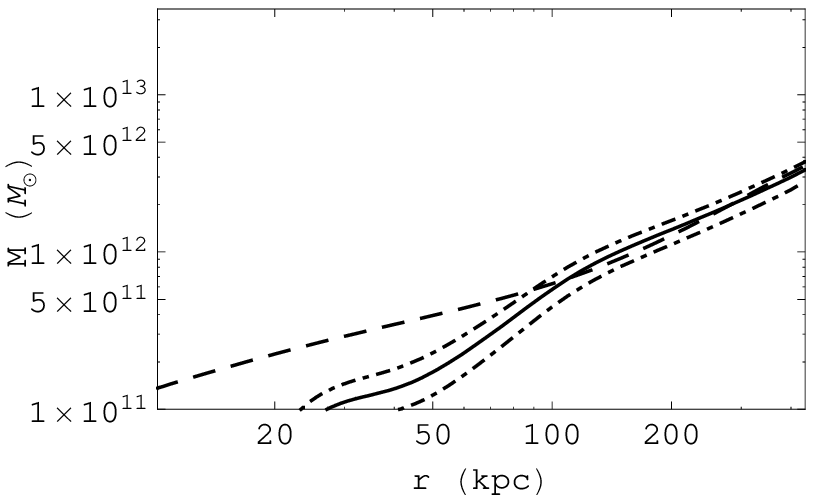}
  \includegraphics[width=85mm]{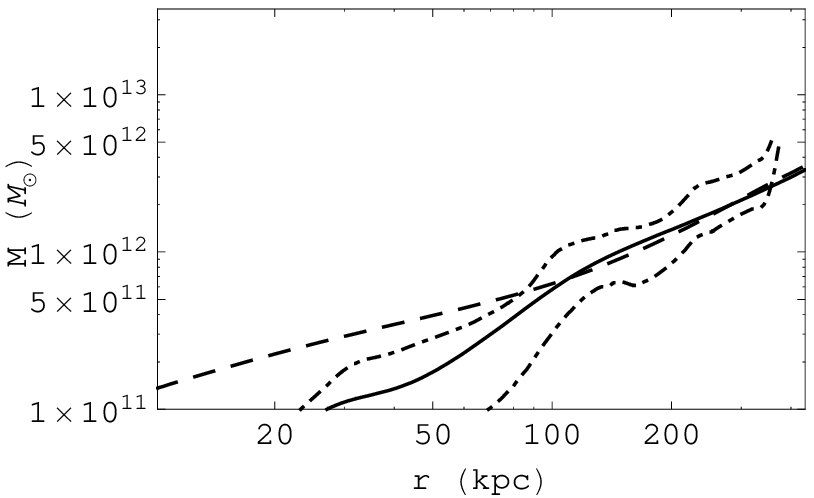}
  \caption{Same of Fig.~\ref{fig:plot_fin_1} but for cluster MKW4.}
\label{fig:plot_fin_11}
\end{figure*}

\begin{figure*}
\centering
  \includegraphics[width=85mm]{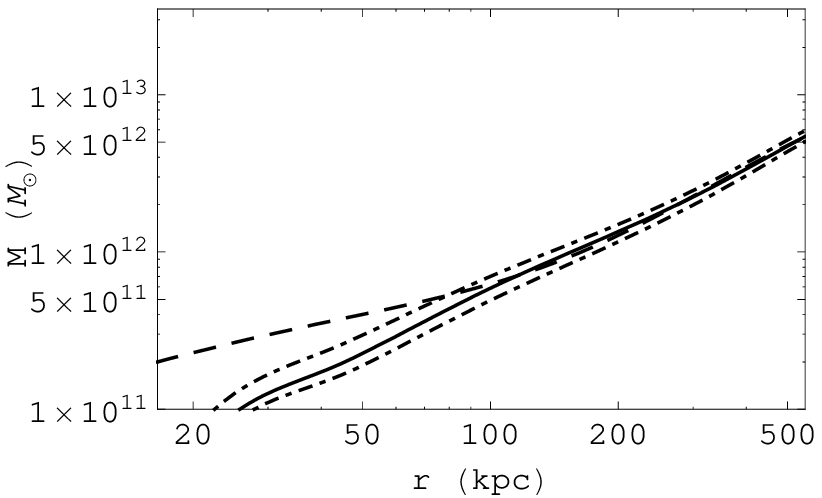}
  \includegraphics[width=85mm]{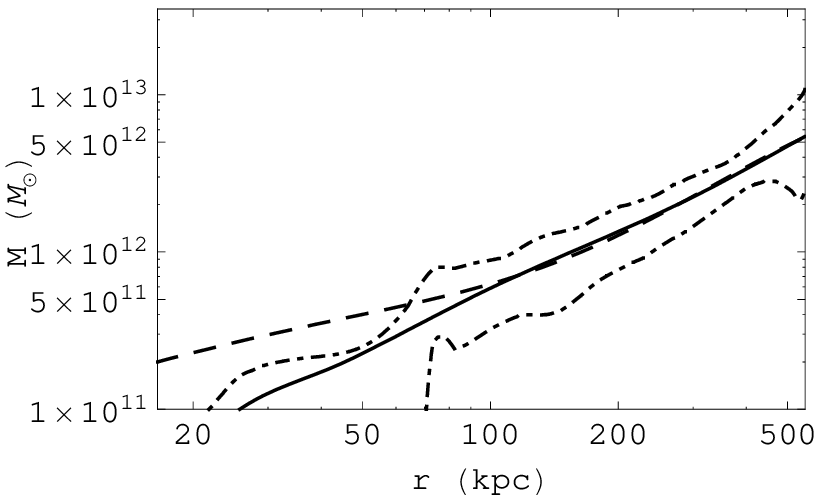}
  \caption{Same of Fig.~\ref{fig:plot_fin_1} but for cluster RXJ1159.}
\label{fig:plot_fin_12}
\end{figure*}

\begin{figure*}
\centering
  \includegraphics[width=84mm,height=60mm]{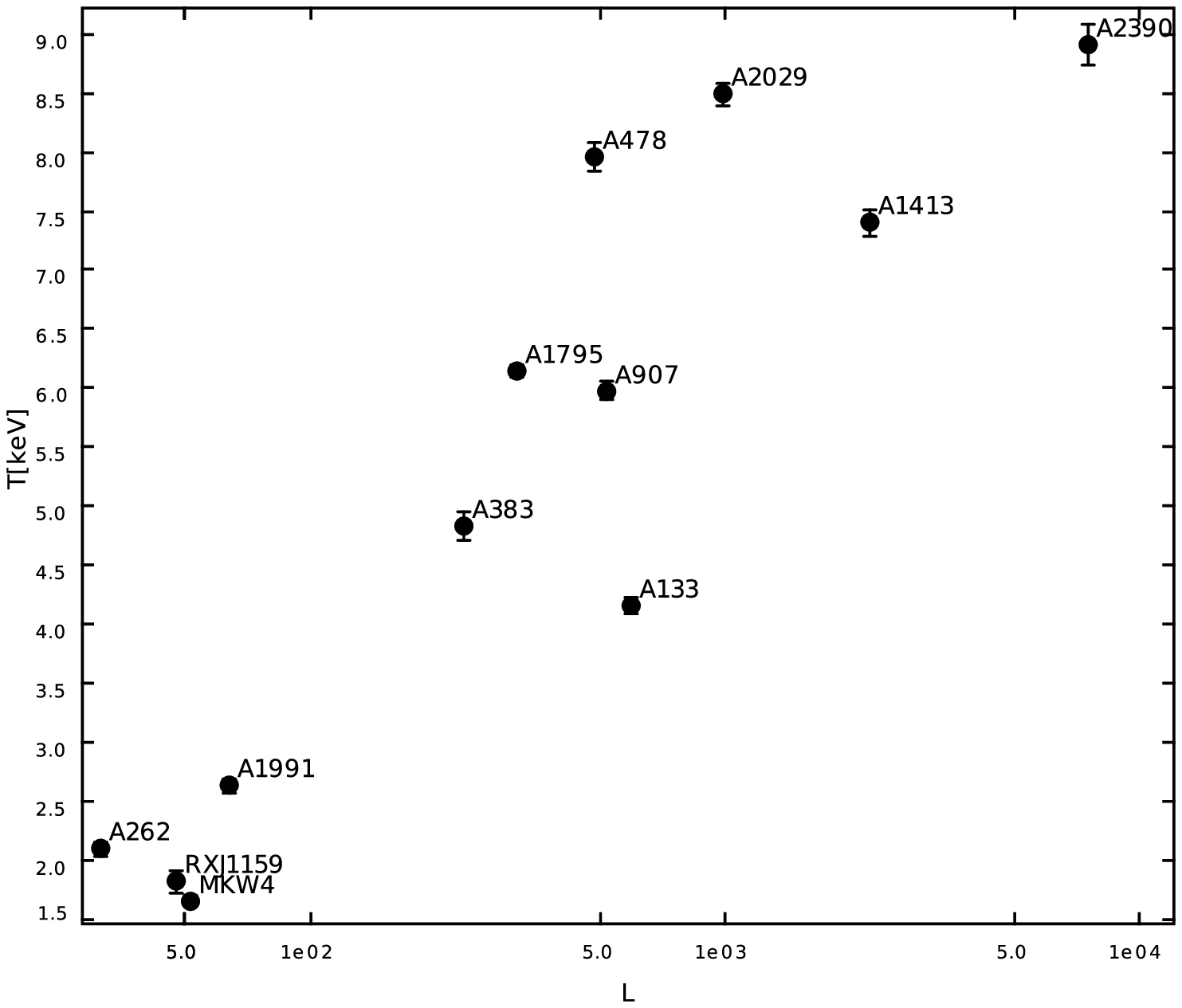}
  \includegraphics[width=84mm]{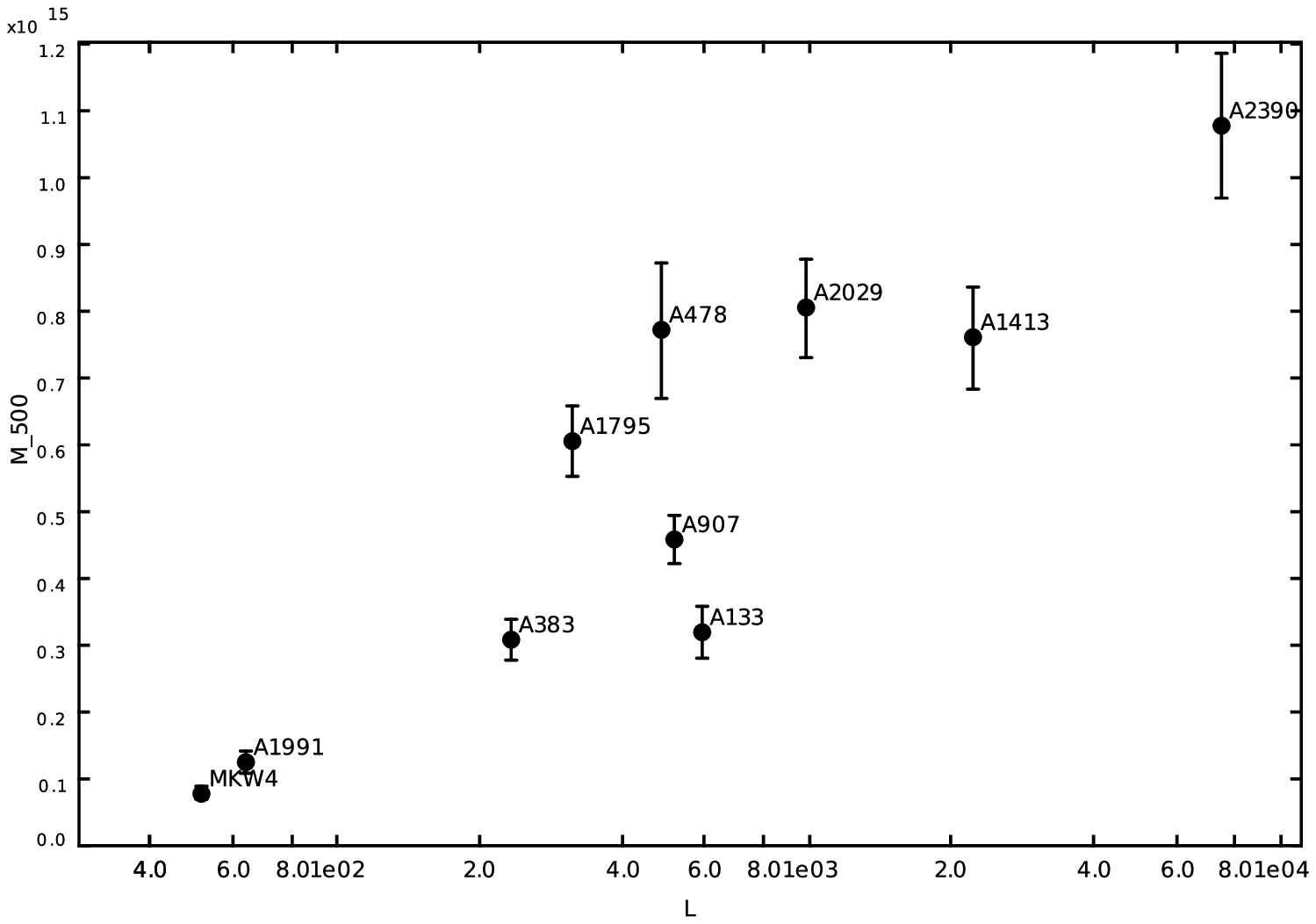}
  \caption{Single temperature fit to the total cluster spectrum (left panel) and total cluster mass within r$_{500}$
  (given as a function of M$_{\odot}$) (right panel) are plotted as a function of the characteristic
  gravitational length $L$. Temperature and mass values are from Vikhlinin et al.2006.}
\label{fig:PropPhysVSL}
\end{figure*}

\label{lastpage}

\end{document}